\documentclass{article}

\usepackage{arxiv}

\usepackage{graphicx}
\usepackage{floatrow}
\newfloatcommand{capbtabbox}{table}[\captop][\FBwidth]
\usepackage{subcaption}
\usepackage{soul}

\floatstyle{plaintop}
\restylefloat{table}

\usepackage{algorithm}
\usepackage[noend]{algpseudocode}
\makeatletter
\def\BState{\State\hskip-\ALG@thistlm}
\makeatother

\usepackage{lscape}
\usepackage{pifont}
\newcommand{\cmark}{\ding{51}}%
\newcommand{\xmark}{\ding{55}}%

\usepackage[utf8]{inputenc} 
\usepackage[T1]{fontenc}    
\usepackage{url}            
\usepackage{booktabs}       
\usepackage{amsfonts}       
\usepackage{nicefrac}       
\usepackage{microtype}      
\usepackage{enumitem}
\usepackage{bm}
\usepackage{cancel}

\usepackage[table,xcdraw]{xcolor}
\usepackage[title]{appendix}
\usepackage{amsmath}
\usepackage{adjustbox}
\usepackage{framed}
\usepackage{float}
\usepackage{natbib}
\usepackage[normalem]{ulem}

\usepackage[pdfencoding=auto,colorlinks,citecolor=blue]{hyperref}

\definecolor{whitesmoke}{rgb}{0.96, 0.96, 0.96}
\definecolor{shadecolor}{named}{whitesmoke}

\definecolor{powderblue(web)}{rgb}{0.69, 0.88, 0.9}
\definecolor{antiquewhite}{rgb}{0.98, 0.92, 0.84}
\definecolor{navajowhite2}{RGB}{ 238,207,161}

\usepackage[colorinlistoftodos,textwidth=1.8cm, textsize=tiny]{todonotes}

\title{\bf \LARGE Assessing competitive balance in the English Premier League for over forty seasons using a stochastic block model}

\author{Francesca Basini$^{1}$ \and Vasiliki Tsouli \and Ioannis Ntzoufras$^2$ \and Nial Friel$^{3,4}$\footnote{Address for correspondence: \texttt{nial.friel@ucd.ie}}}
\date{\small
    $^1$Department of Mathematics, University of Warwick, UK\\%
    $^2$ Department of Statistics, Athens University of Economics and Business, Greece\\
    $^3$School of Mathematics and Statistics, University College Dublin, Dublin 4, Ireland\\
    $^4$ Insight Centre for Data Analytics, University College Dublin, Dublin 4,Ireland\\[2ex]%
    \today
}
\begin{document}
\maketitle

\begin{abstract}
Competitive balance is the subject of much interest in the sports analytics literature and beyond.
In this paper, we develop a statistical network model based on an extension of the stochastic block model to assess the balance between teams in a league. Here we represent the outcome of all matches in a football season as a dense network with nodes identified by teams and categorical edges representing the outcome of each game as a win, draw or a loss.
The main focus and motivation for this paper is to provide a statistical framework to assess the issue of competitive balance in the context of the English First Division / Premier League over more than $40$ seasons.  The Premier League is arguably one of the most popular leagues in the world, in terms of its global reach and the revenue which it generates. Therefore it is of wide interest to assess its competitiveness. Our analysis provides evidence suggesting a structural change around the early 2000's from a reasonably balanced league to a two-tier league.
\end{abstract}


\section{Introduction}

Uncertainty in the outcome in sporting events would, on the face of it, appear to be an essential ingredient of attractive team sports \citep{Szymanski2001}.  
The study of competitive balance is one which has interested researchers for some time, particularly in fields as diverse as economics to sports science. For example, there is much interest to investigate relationships between the competitiveness of a sporting league and the distribution of revenue within the league. For example, \cite{brandes07}
study the effect which competitiveness of the Bundesliga in Germany has on attendance 
while \cite{manasis_etal_2021} study the effect of competitiveness for several European leagues; 
see also \citep{penn2014} and \citep{plumley2018} who present a perspective in the context of the English Premier League.
More generally, competitive balance is of vital interest and used as an argument by sports governing bodies to ensure that, for example, revenue streams are allocated proportionately. In fact, in the context of soccer, that was partly the reason for introducing the UEFA Financial Fair Play Regulations (FFP) to limit excessive spending and investments by club owners together with the intention to ensure that clubs budgets are regulated, reducing debts from season to season and creating profit. The new regulations were agreed in 2009 and put in practice three years later.

In the literature of sports analytics, several authors have proposed statistical measures of competitive balance. 
Most of these indices are based on intuition about the notion of competitiveness.
These are typically descriptive statistics based on summarising the spread of points or wins achieved by each team at the end of the league, see \citep{Evans_2014} and references therein. For example, the Herfindahl--Hirschman index of competitive balance is adapted from the Herfindahl-Hirschman index which is used to measure the spread of market share by firms in a given industry. Other definitions are based on the idea of entropy among many others. We refer the reader to \citep{manasis2013quantification}
for a statistical perspective on this topic; see also \citep{Evans_2014} for a comprehensive review of this literature. 
A thorough search of the related methodology reveals very few publications with competitive balance indexes which are based on solid statistical models or techniques. 
Some exceptions include the approach of \cite{Koning_2000} based on multinomial probit type of model with response the game outcome (win, draw, loss), the index introduced by \cite{Haan_etal} based on a regression model for the goal difference, and the autoregressive win percentage by \cite{Vrooman_1995} which is based on an auto-regressive model. 
Further, model-based approaches are the variance decomposition measure \citep{Eckard_1998} and CBR \citep{Humphreys_2002}, which are  based on ANOVA type models. 
Finally, two applications of Markov models provided measures of competitive balance of
sports leagues: (a) the statistical test of theoretical and actual transitional probabilities which allows for the testing of a wide range of hypothesis regarding competitive balance relating to strata of a league structure 
\citep{Koop_2004} and 
(b) a ‘Gini type’ single statistic measure of the competitive balance of a league system \citep{Buzzacchi_etal_2003}.

The notion of competitiveness in sports is multidimensional with many qualitative characteristics. This has generated a considerable number of different indexes and measures of competitiveness. 
One of the characteristics, well explained in the Economics literature is the so-called  three-dimensional factorization of competitive balance; see for example \citep{cairns_1987}.
These three dimensions are: a) the match uncertainty, which refers to a particular game, b) the seasonal dimension, which focuses on the relative quality of teams in the course of a particular season and c) the between-seasons dimension, which focuses on the relative quality of teams across seasons. 
Another aspect that has recently been considered by \cite{manasis2013quantification} is the multi-levelled nature of European leagues offering multiple awards  as opposed to the common single prize offered by North American ones. 
These multiple awards are related with qualification positions (of different quality) for playing at UEFA Champions League, Europa League and to league positions leading to relegation. 
The attention of the economic analysis of competitive balance is the effect on the fans’ behaviour, which is the longstanding ``Uncertainty of Outcome Hypothesis'' 
\citep{Fort_Maxcy_2003}.

The statistical model which we develop is based on the idea of a stochastic block model (SBM) which is popular in the analysis of relational network data \citep{nowicki2001estimation}. The central idea of an SBM is to partition the nodes of a network (or graph) into blocks (or clusters) so that every pair of nodes within a block tends to have the same probability of being connected by an edge and that this connection probability varies by block. 
But also, the connection probability of any two nodes in different blocks depends on the blocks to which each node belongs.
We extend this framework to data arising from a football league. By analogy, we conceptually consider teams as being nodes of a complete graph, where every pair of nodes is connected by an edge and also that the relational edge between two teams is the categorical outcome, a win, draw or loss, resulting from when they play against each other. In particular and by analogy to a standard SBM, we aim to partition teams into blocks so that the outcome (a win, draw or loss) when two teams from a same block play each other follows the same multinomial distribution, the parameters of which vary according to the block label of each team. Effectively, teams within a block are considered to be \textit{balanced} as the outcome for each game follows the same multinomial probability mass function. 
Additionally, the SBM framework also models edges involving teams in different blocks with a multinomial probability mass function which depends on the blocks to which each team belongs. Effectively, the assumption that the multinomial probability mass function for any pair of nodes depends on the block to which each team belongs to incapsulates the concept of \textit{stochastic equivalence}, first introduced by \cite{nowicki2001estimation}.
As such, an important aspect of our modelling framework is to infer both the number of blocks, but also to assign probability to the membership or allocation of each team to a block. 
In particular, if we infer that a single block is most likely, this suggests that the league is balanced. While additionally estimating the number of teams in the strongest block, integrating over the posterior uncertainty in the number of blocks, provides a further means to quantify  competitiveness. For example, if we find that a two block model has most support, but that the weakest block contains only two teams, then there is evidence that the league is quite competitive for the teams in the top block.

Hence, in this paper we offer an innovative implementation of an SBM for football (which can be used also for other competitive sports between two opponents). 
First of all, we should emphasize that the proposed method relies on more granular data than other methods in the literature which only use the information from the final league table.
Importantly, relational information of each individual game is considered instead. By considering the game specific outcome, we directly consider competitiveness in multiple dimensions: for each game separately and in terms of seasonal dimension. Moreover, the across season competitiveness can be also evaluated by analysing the main outcomes of SBM across different seasons. 
Our proposed method is solely based on a solid statistical model and not on intuition leading to multiple outcomes for measuring different characteristics of competitive balance. 
To be more specific: 
\begin{enumerate}
	\item Primarily, the number of groups specify whether the teams can be classified in different blocks/levels of competitiveness. This can be also related with the different levels of awards offered in a league. An SBM consisting of one group correspond to leagues where all teams have similar probabilities to win against each other, implying a balanced league. 
	\item The number of teams belonging in the top block/cluster is also an important index that can be evaluated in terms of competitiveness. Even in cases of two groups, a top-team block with many members indicate high competitiveness for the higher league positions.
	Note that, as \cite{manasis2013quantification} have indicated, the competitiveness for top league positions is more important for the fans than competition for bottom league positions (leading to relegation). 
	\item The persistence of specific teams appearing in the top-teams group is an important measure of across-seasons competitiveness.
\end{enumerate}
Note that our proposed SBM method evaluates the overall competitiveness of a league and not the competition for the overall  champion each season, which is very important from the point of view of fans; 
see \citep{manasis2013quantification} for a related discussion and empirical evidence.

The motivation for this paper was specifically to analyse the top league in English football, namely the First Division / Premier League, to try to shed light on the question of whether this league has become more imbalanced over time. 
The English football league is one of the oldest football leagues in the world dating back to $1888$ when the league consisted of only $12$ clubs. It grew rapidly with the introduction of a second division in $1892$ and today consists of $4$ divisions. 
Interestingly, no club has been ever present in top division. Everton holds the record of most seasons in the top division, missing only $4$ seasons in total, while Arsenal and Aston Villa have both amassed over $100$ seasons. A persistent feature of the English football leagues is that of relegation from the First Division and promotion from the Second Division; similarly, for the other lower divisions. In the more recent past, encompassing the study period of this paper, the English First Division has undergone several changes in its structure and format. We outline these briefly here. In $1978/79$, at the beginning of our study period, the First Division consisted of $22$ teams. This was reduced to $21$ teams for a single season in $1987/88$ before reducing further to $20$ teams until $1990/91$. It then reverted to $22$ teams from $1991/92$ to $1994/95$, before changing once again in $1995/96$ to $20$ teams where it has remained to date. Another important change which occurred over the period of study in this paper was the adoption of \textit{3-points-for-a-win} in $1981/82$. The motivation for this change was to encourage more attacking playing tactics than the previous \textit{2-points-for-a-win} as it was widely accepted that teams would often settle for a draw since the difference was only one point compared to a win. We note that this change occurred towards the beginning of our study, so this might have minimal effect on the conclusions which we make. Although on the other hand, our model does not account for the number of points won, so may be agnostic to this change in any case. However, perhaps the most significant change has been the introduction of the English Premier League in $1992/93$. This heralded massive increases in revenue primarily through satellite television payments to clubs. The first TV deal between the Premier League and the television companies generated revenue of around $40$ million pounds. This has increased dramatically over time to $5.14$ billion between $2016$ and $2019$ \citep{ernstyoung17}. As such it is not an understatement to say that the introduction of the Premier League has had a transformative effect on football. 

There is speculation in the literature that competitive balance is associated positively with increased fans' interest eventually leading to increased income or ticket sales; see for example \citep{manasis_etal_2021}. 
Nevertheless, the results of empirical studies on the relationship between competitive balance and audience demand are mixed. 
In a large number of such studies, the evidence of a positive association between match uncertainty and demand is either relatively weak or even contradictory; see \citep{Borland}, \citep{Pawlowski} and \citep{Coates_etal}. 
While, evidence concerning the existence of an effect of the seasonal and between-seasons competitive balance measures on demand is more systematic; 
see, for example, \citep{Humphreys}, and  \citep{manasis_etal_2021}.

The perceived wisdom appears to be that the First Division / Premier league has gradually become more imbalanced especially since the inception of the Premier League in $1991$. Moreover, it is commonplace to see discussion in the media regarding the \textit{big-six} teams, usually referring to Arsenal, Chelsea, Liverpool, Manchester City, Manchester United and Tottenham Hotspur, as having emerged in the recent past. The reason for this is usually explained by the fact that there is strong competition for the first six places as the teams in these positions qualify for the European competitions, notably the UEFA Champions League, which in turn generates more income for these teams. 
Yet, it is not immediately clear how grounded in reality this notion of a \textit{big-six} actually is. 
To the best of our knowledge, there has been no statistical modelling framework in place to answer these and other questions. Our aim is to fill this gap.

To address these questions, we apply our model to over $40$ seasons of the old English First Division / Premier league. Our findings are that the league was relatively balanced until the early part of the $2000$s,
but that it has become quite imbalanced since then. Over the first half of our study from $1978/79$ to around $2002/03$ we find many seasons for which a single block model is preferred and that in seasons where a two block is estimated to have highest posterior probability, that there is often considerable support for a single block model. Additionally, we find that the number of teams in the strongest block is typically large, again suggesting that the league was relatively balanced for the majority of seasons prior to $2000$. This is in contrast with our findings for the second half of the study period. 
In particular, we find that since the $2003/04$ season a two block model is typically best supported by the data and that membership to each block has become quite sharp or polarised, in the sense that the probability of allocation of any team to the strongest block is either very high or very low. Moreover, since $2003/04$ the number of teams in the top block varies between $2$ and $8$ teams (with the exception of only three seasons, $2010/11$, $2015/16$ and $2020/21$).

This paper is structured as follows. We begin in Section~\ref{sec:comp_balance} by providing a brief summary of two measures of competitive balance and indicate their shortcoming. We present the form of the data which we analyse in Section~\ref{sec:data} pointing out that the data can be considered as an adjacency matrix of a complete network. Section~\ref{sec:model} develops the stochastic block model for this type of data. While Section~\ref{sec:mcmc} outlines an MCMC algorithm which we use to sample from our Bayesian model. The substantive application of our model to $42$ seasons spanning the old English First Division through to the present day Premier League is presented in Section~\ref{sec:results}. The paper concludes with Section~\ref{sec:conclusion} and outlines some potential extensions of this framework to incorporate a statistical changepoint analysis and also temporal dependence between seasons.

\section{Statistical approaches to assess competitive balance}
\label{sec:comp_balance}

The notion of competitive balance is complicated, having multiple different perspectives commonly referred as dimensions or factors. 
In fact, there has been considerable debate in the relevant literature about the nature of competitive balance in football or other sports over recent decades.
For this reason, many different measures of competitive balance or imbalance have been introduced in the bibliography trying to capture these different characteristics;  see, for example,
\citep{Vrooman_1996}, \citep{Zimbalist}, \citep{Fort}, \citep{Szymanski} and \citep{Pawlowski_Nalbantis}. 
One popular notion in competitive balance is the three-dimensional factorization 
\citep[see, for example, ][]{Quirk_Fort,Czarnitzki_Stadtmann,Borland,brandes07} 
which is composed of a) the match uncertainty, b) the within season or seasonal uncertainty, and c) the between-seasons uncertainty.

A wide range of measures have been introduced which have been designed to
capture the degree and the different characteristics of competitive balance and imbalance. 
Each of these measures has advantages and disadvantages which try to encapsulate a complicated notion with a single summary index. 
In the end, there is no single optimal measure of competitive balance which is widely accepted as ``correct'' or most appropriate for each problem, league or sport. Indeed, each competitive balance measure focuses on a different feature of the problem.

Within season uncertainty is focused on measuring the differences in the relative quality or strength of teams during a particular season. 
Some selected seasonal competitive balance indices include 
the National Measure of Seasonal Imbalance \citep{Goossens}
the Herfindahl-Hirschman Index \citep{Depken}
and its standardized version \citep{Owen_2007}, 
the Adjusted Gini Coefficient \citep{Schmidt, Utt_Fort}, and 
the specially designed indices introduced by \cite{manasis2013quantification}. 
Between-seasons indices focus on the ranking differences of the final result between subsequent seasons and they include measures such as 
the index of Dynamics of \cite{Haan_etal}, and
the special indices of \cite{Manasis_Ntzoufras_2014}. 
For a more thorough review and detailed discussion concerning the applicable indices to European football see  
 \citep{Manasis_Ntzoufras_2014}.

Here we focus on two indicative popular indices used to measure specific aspects of competitiveness in sports leagues and then we proceed with our proposed method.  
The first index is the Herfindahl--Hirschman index (HHI) which was originally used to measure the dominance of a firm or company to a specific industrial field and, hence, it is an indicator of the amount of competition among them \cite{Hirschman_1945,Herfindahl_1950,Hirschman_1964}. 
HHI naturally lends itself to the context of competitive balance in sports \citep{Owen_2007} since some of the basic notions are in common with company dominance in industry. 
Indeed we note that the framework which we subsequently develop may in turn find use in terms of measuring company dominance in settings where relational data involving interactions between companies is present, for example trade flows from one company to another, noting that extensions to numerical edges would be relevant in this context.
The second measure we explore is relative entropy which is an asymmetrical measure of similarity between the relative frequencies of the data and a theoretical distribution. Here the relative frequencies are replaced by the proportion of points won by each team and this is compared with the proportion of points expected in a perfectly balanced league. 
Both of these measures are representative of a group of competitive balance measures that focus on the spread of points or wins at the end of the season. Hence all the relevant information is taken from the final league table; see 
\citep{manasis2013quantification} and \citep{Evans_2014} for a comprehensive review of relative measures. Note that the relative entropy can be also adopted to represent game-by-game uncertainty but, to the best of our knowledge, it has not been used in this context in the competitive balance literature. 

Here, we introduce a sophisticated statistical 
approach based on network modelling in order to analyse the competitiveness between teams in a football league. 
Our proposed method accounts for individual games and automatically identifies groups/clusters of different strength. Single block leagues are indicative of high competitiveness where every team can beat every opponent. 
This is, intuitively, a seminal characteristic of the Premier league for specific seasons according to the perspective of football fans and sports media. On the other hand, two block leagues may indicate a two-stage competition between teams of different strength (but not always -- this depends on which and how many teams are separated from the main body of the league and the posterior uncertainty in the number of blocks).

\subsection{Herfindahl--Hirschman index of competitive balance}

Perhaps the most widely used means to assess competitive balance within a season in a sporting league is the Herfindahl--Hirschman index of competitive balance \citep{Owen_2007}. This index is based on assessing a measure of the spread of points share in a given season. Suppose that team $i$ scored $s_i$ points over the course of a season in a league involving $n$ teams. Then one defines the HHICB index of competitive balance (HHICB) in terms of $p_i := s_i/\sum_1^n s_i$, the proportion of points achieved by team $i$, as
\[
 \mbox{HHICB} = n \sum_{i=1}^n p_i^2. 
\]
It is therefore simply a measure of the variability of the vector $(p_1,\dots,p_n)$. When each team has an identical proportion of points so that $p_i=1/n$, then $\mbox{HHICB=1}$.
While the higher the value of NNICB, the more imblanced the league is. In Figure~\ref{fig_rel_entropy} (a) the $\mbox{HHICB}$ statistic is plotted for each season from $1978/79$ to $2021/22$, the duration of the study in this paper. We remark that there is a general trend that the $\mbox{HHICB}$ statistic is increasing over time lending some evidence to the hypothesis that the league has become more imbalanced over time.

\subsection{Relative entropy as a measure of competitive balance}

\begin{figure}[h]
 \centering
 \begin{tabular}{cc}
\includegraphics[scale=0.3]{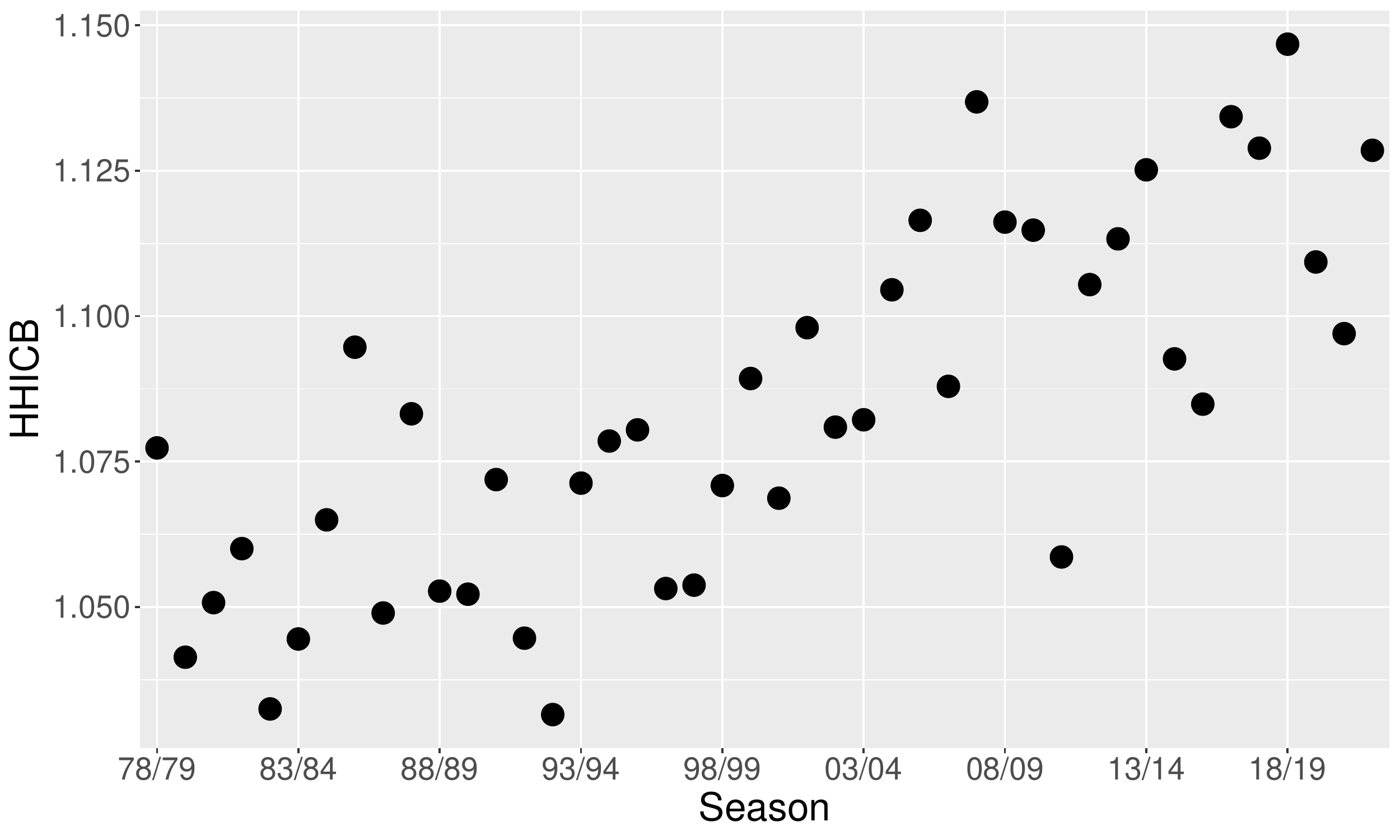}
&
\includegraphics[scale=0.3]{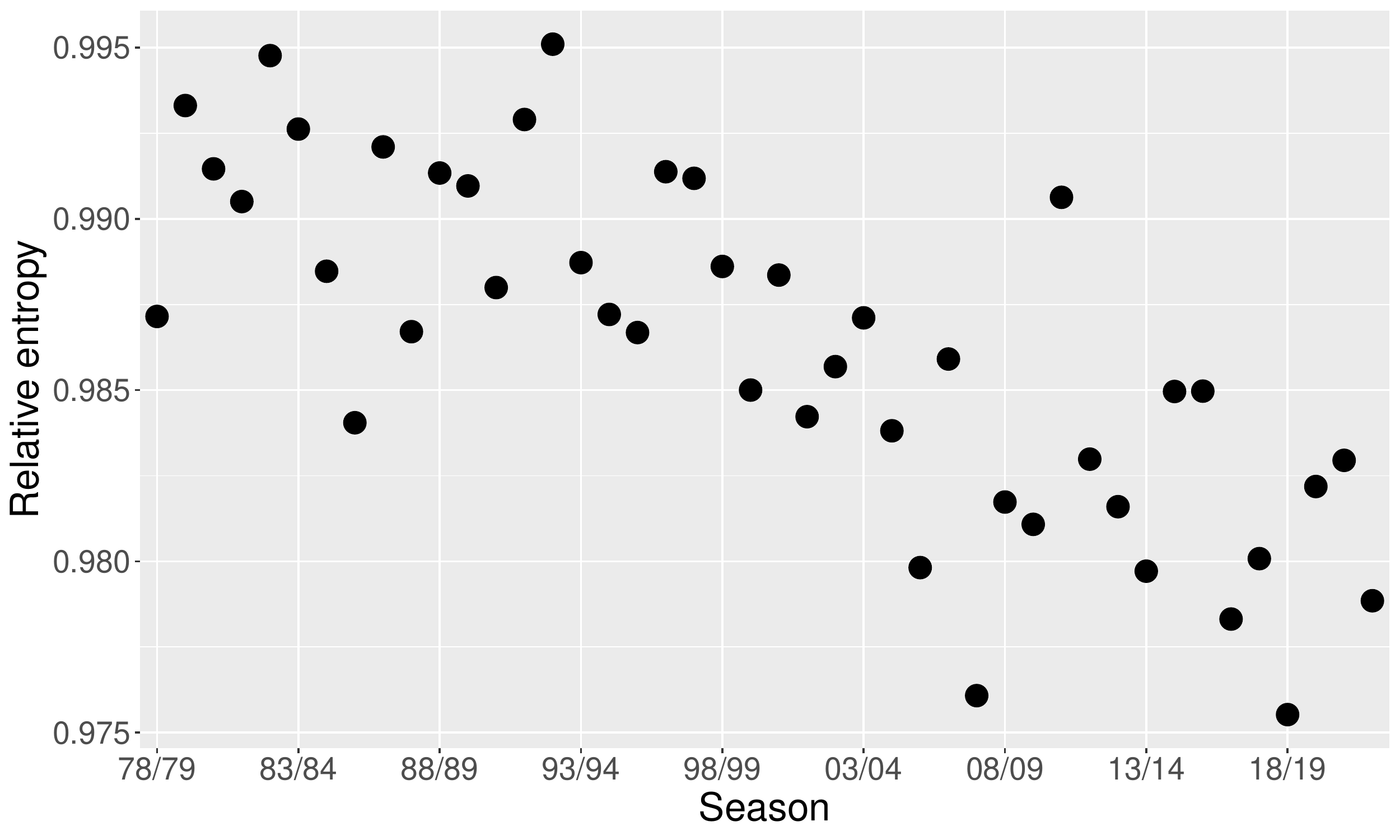}
\\
(a) & (b)
\end{tabular}
\caption{(a) The Herfindahl--Hirschman index of competitive balance (HHICB) is plotted for each season. This illustrates a general trend that HHICB is increasing over time and so consistent with the notion that the Premier League has become more imbalanced over time.
(b) Relative entropy is plotted for each season. Here high values of relative entropy correspond to more balanced league and this plot suggests that the Premier League is gradually becoming more imbalanced over time.}
\label{fig_rel_entropy}
\end{figure}

A natural approach to summarise the vector of proportion of points share among all $n$ teams in a league, $(p_1,\dots,p_n)$, is to use the concept of entropy \citep{horowitz97}. Here one may define the relative entropy for a given season as: 
\[
 \frac{\sum_{i=1}^n p_i \log(p_i)}{\log(1/n)}. 
\]
This statistic takes a maximum value of $1$ in the case where $p_i$, the proportion of points share for team $i$ is $1/n$ for all $n$ teams and this corresponds to the case of a perfectly balanced league. While lower values of relative entropy correspond to a more imbalanced league. Figure~\ref{fig_rel_entropy} (b) displays the relative entropy statistic for each season over the course of this study. Here there is a general trend towards lower relative entropy over time, once again coherent with the hypothesis that Premier league has generally become more imbalanced over time.

\subsection{The proposed statistical modelling approach}
\label{sec:alt_model}

Each of the two measures of competitive balance outlined in the previous sections are somewhat limited as  it is not straightforward to make any qualitative conclusions such as whether a season is balanced. Moreover, if there is evidence to suggest that a season is imbalanced, it would be useful to give an indication as to the nature of this imbalance, for example, which teams, if any, are stronger than the rest. To date, to the best of our knowledge, there is no literature which provides such a statistical framework. We aim to address this shortcoming here and develop a fully probabilistic model of competitive balance using a stochastic block modelling framework.
The proposed network model offers  multiple outcomes with information about the quality of the leagues and the relative competitiveness. 
Generally, the method we propose is more computationally demanding but richer in terms of results and inference we obtain from the final output it offers. 
Essentially the idea, which we now develop, is based on a statistical model which probabilistically partitions teams into blocks (or clusters) where the outcome (a win, draw or loss) for the home team follows the same multinomial probability mass function when any two teams in the same block play one another.
Similarly, the outcome when any two teams which belong to different blocks play one another follows a multinomial probability mass function whose parameters depend on the blocks to which each team belong.
In this way if we estimate that a single cluster (consisting of all teams) is most probable, then this provides evidence that the league is balanced. Moreover, in the case where the model estimates support for a league with more than one cluster, we assign probabilities of membership (or allocation) for each team to any of the blocks. This in turn would allow one to assess the presence of blocks or clusters of teams which are broadly competitive with each other. Additionally, by integrating over the uncertainty in the number of blocks we can estimate the number of teams in the strongest block and this provides another means to assess competitive balance. This perspective is particularly important for seasons in which there is broadly equal support for a one or a two block model, hence accounting for this uncertainty is an important factor to accommodate. 
Assessing the constituent teams in the strongest block of teams over time is of interest in the context of the English Premier League as there is anecdotal evidence of the emergence of a \textit{big-six} block of teams over the past decade or so, namely, a collection of teams which are stronger than the remaining teams leading to a competitively imbalanced league. Our work aims to provide Bayesian framework to tackle these types of questions.

\section{Representing the outcome for a season as a results matrix}
\label{sec:data}

\begin{figure}[H]
	\begin{tabular}[h]{cc}
		\centering
		\includegraphics[scale=0.3]{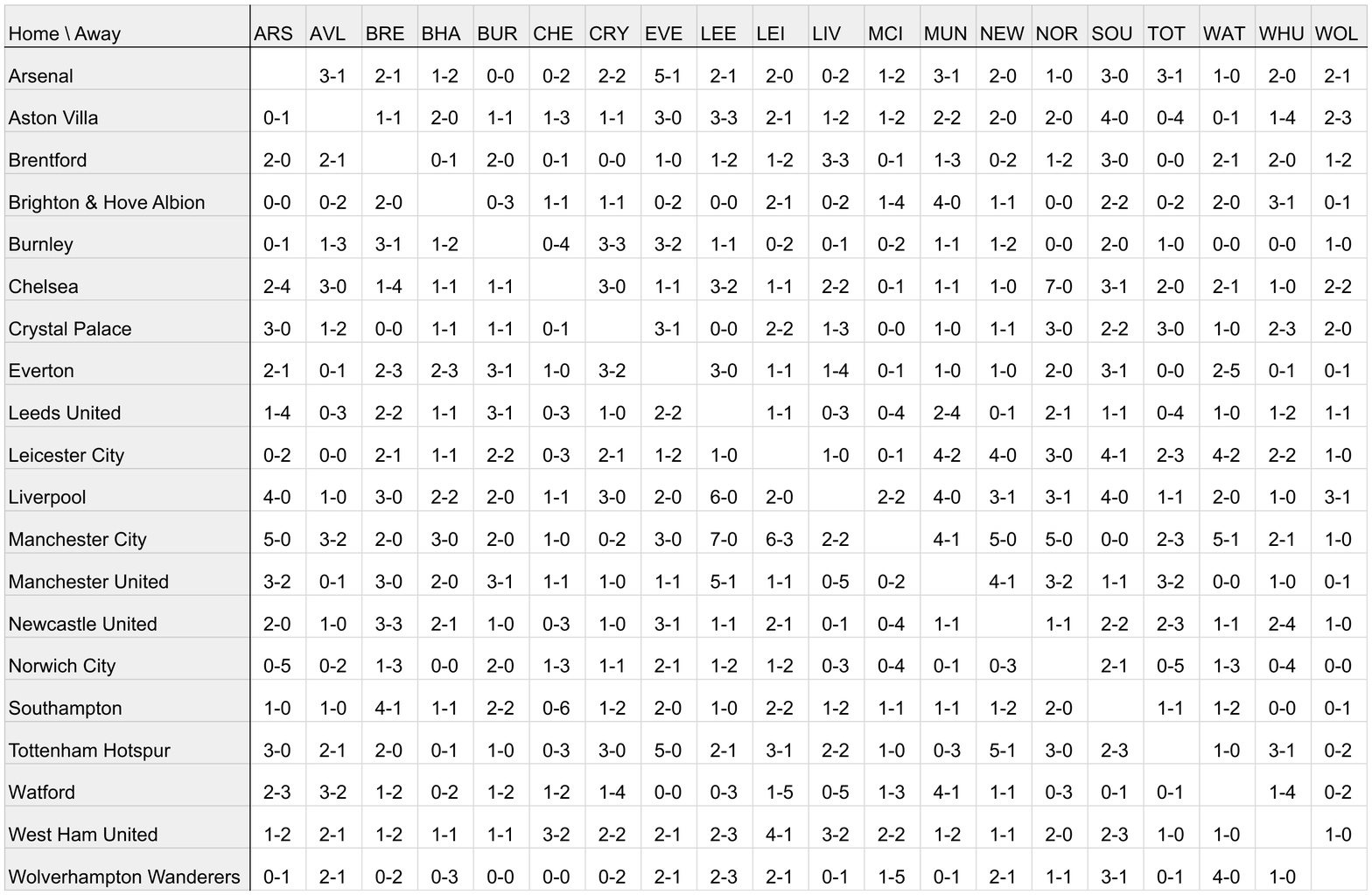} &
		\includegraphics[scale=0.35]{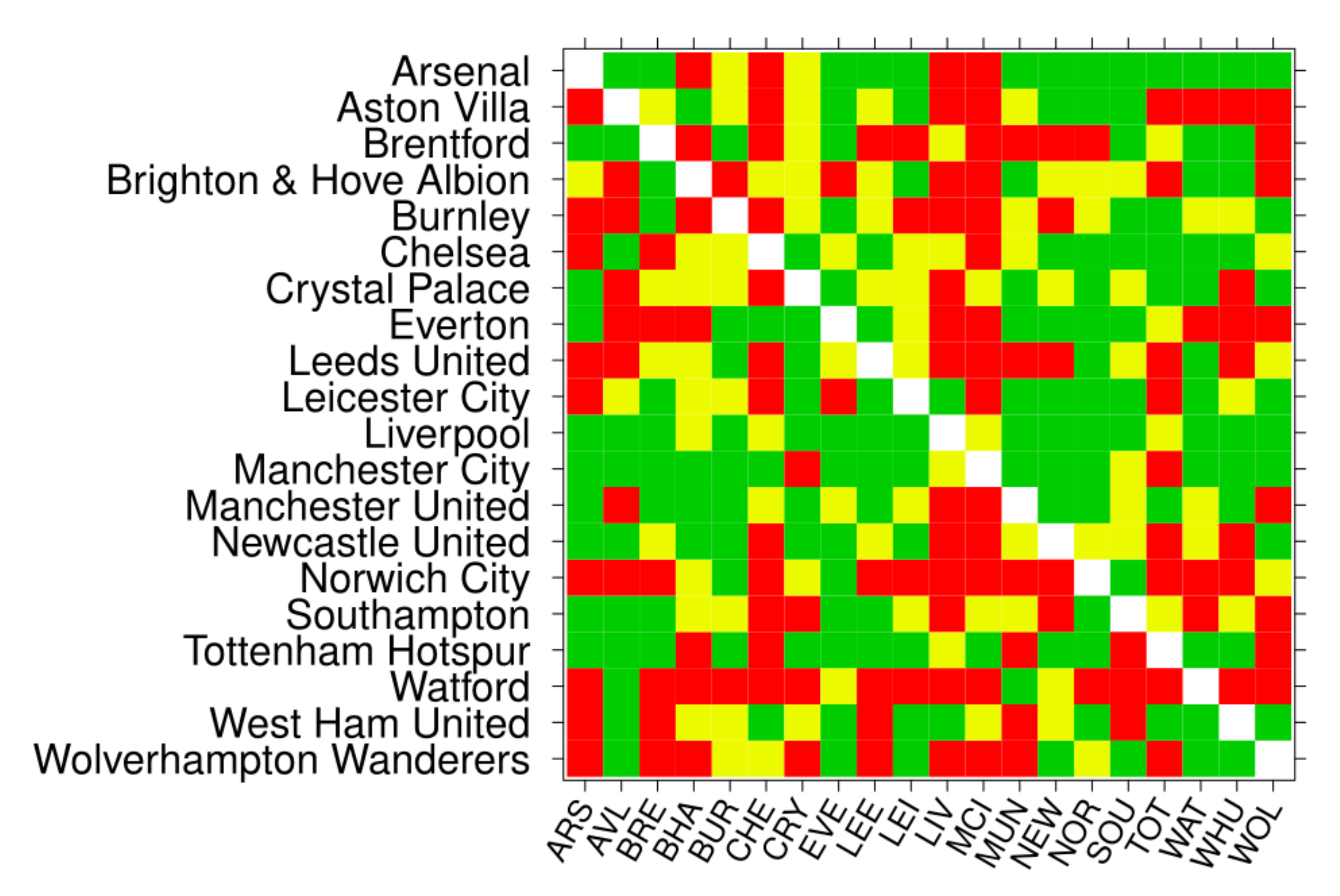} \\
		(a) & (b) \\
	\end{tabular}
	\caption{The matrix of results of the 2021/22 Premier League season. (a) Cell entries correspond to the result when a home team (row) plays an away team (column). 
		(b) The results in (a) summarised in a results matrix by categorising each result as a win, draw or loss corresponding to the colour green, yellow or red, respectively. }
	\label{fig:MatchGrid_2122}
\end{figure}

To begin, we introduce the format of the data which we analyse each season. In particular, we represent the entire collection of results for every game in a single season in the form of a matrix. Here we consider the typical league scenario where each team plays every other team twice, once at home and once away from home. Therefore in a league with $N$ teams, there are $N\times(N-1)$ fixtures. The results of all these fixtures can be summarised in a $N\mathsf{x}N$ matrix, $R$. Each cell $r_{ij}$ contains the result of team $i$ playing against team $j$ when team $i$ plays at \textit{home}. Clearly, the diagonal entries are missing as no team plays against itself. The matrix for the season $2021/22$ is displayed in Figure~\ref{fig:MatchGrid_2122} (a).

In the modelling framework considered in this paper we take as outcome variable for every game, the three categorical variables, "win", "draw" or "loss" for the home team, denoted by $1,2$ or $3$, respectively. Therefore this leads us to summarise the matrix $R$ with a transformed matrix $\bm{y}$ which we term a results matrix, 
\begin{equation}
 \bm{y}= 
\begin{pmatrix}
			- & y_{12}  & \dots  & y_{1j}&\dots &y_{1N} \\
			y_{21} & - &  \dots  & y_{2j}&\dots&y_{2N} \\
			\dots & \dots & - & \dots & \dots \\
			y_{i1} & y_{i1} & \dots & -& \dots & y_{iN}\\
			\dots &\dots&\dots&\dots&-&\dots\\
			y_{N1} & y_{N2}  & \dots &  y_{Nj} & \dots &-
		\end{pmatrix}\, ,  
\label{eqn:res_matrix}
\end{equation}
where $y_{ij}\in\{1,2,3\}$, for $i=1,\dots,N; \quad j=1, \dots , N; j\neq i.$ The categorical outcomes for season $2021/22$ are displayed in Figure~\ref{fig:MatchGrid_2122} (b), where the categorical variables "win", "draw" or "loss" are displayed using the colours green, yellow and red, respectively.

We remark that the results matrix $\bm{y}$ can be considered as an adjacency matrix of a directed network where each node represents a team and the outcome of a match between home team $i$ and away team $j$ is represented by the weight of the edge $y_{ij}$, where the weight takes a value in the set $\{1,2,3\}$. In many networks, we observe sparsity in the relational variable. For example, in a social network where the dyadic relational variable is binary, for instance a friendship relationship recorded as $1$ if the two nodes are friends, and $0$ otherwise, typically most dyads will take the value $0$. However this is not the case here as we always record a categorical outcome between each pair of nodes in the network. Hence, the network under study is dense and complete but does not allow self-loops.

\section{A modification of the stochastic block model}
\label{sec:model}

The stochastic block model (SBM) of \cite{nowicki2001estimation} has proven to be incredibly useful in the analysis of relational network data. The core idea of an SBM, which we make precise in the subsequent section, in the context of a binary network is to partition the nodes of the network into blocks such that the probability of an edge between any two nodes depends on an unobserved (or latent) variable assigning each node to one of $K$ blocks or clusters in the network.
One of the contributions of this paper is to explain how this framework can be extended to the context of football results data by developing a Bayesian stochastic block model which can be applied to the results matrix illustrated in the previous section.

Note that the model which we develop falls within the general framework of \cite{aicher2015} where a generalisation of an SBM is presented for edge weights drawn from a exponential family distribution.
We also note that in common with the generative weighted SBM approach of \cite{aicher2015} we consider the case where the observed network is fully observed and is dense. That is to say that all possible interactions between each pair of teams are observed as each season each pair of teams plays each other twice, at home and away.

\subsection{Specification of the block model}

As outlined before, the aim of our stochastic block model is to partition the $N$ teams in a league, into $K$ blocks in such a way that the probability of a win, draw or loss for the home team is, broadly speaking, similar when any two teams in the same block play against one and other. 
But equally, that the probability outcome between teams belonging to different blocks, that is, where any team from one block plays at home against another team from a different block also has a similar probability of a win, draw or a loss. Essentially we wish to model the following scenarios. Matches involving teams within a block tend to have similar outcomes, while matches between teams from different blocks tend also to have similar outcomes. Crucially, the probability of a given outcome depends on the blocks to which each team is assigned. One of the key objectives is to infer the most likely value of $K$. In particular, if we deem that $K=1$ has most support, then we have some evidence that the league is balanced. 

Here we develop the model and introduce some notation. We interchangeably use the terms nodes and teams, as from before there is an analogy between $\bm{y}$, the results matrix (\ref{eqn:res_matrix}), and an adjacency matrix of a network. We make two main assumptions:
\begin{itemize}
		\item For a $K$ block (or cluster) model, each node (or team), $i$ for $i=1,\dots,N$, belongs to one of the blocks with membership or allocation label, 
		$z_i \in\{1,\dots,K\}$.
		\item The distribution of the relational structure $\bm{y}=(\bm{y_{ij}})_{1\leq i\neq j\leq N}$ is assumed to be conditionally independent given the latent variable of cluster memberships, $\bm{z}:=(z_1,\dots,z_N)$.
\end{itemize} 
We now describe the various elements of our model. 

\paragraph{Distribution for the allocation vector $\bm{z}$:}
We assume that the entries of $\bm{z}$ are independent and identically distributed following a multinomial distribution:
\begin{equation}
    z_i|\bm{\theta},K \stackrel{iid}{\sim} Multi(1, \bm{\theta}=(\theta_1, \theta_2, \ldots, \theta_K)), \ \text{for} \ i=1,\ldots, N,
\end{equation}
where $P(z_i=k|\bm{\theta},K)=\theta_k $ is the probability that node $i$ belongs to cluster $k$, $\theta_k >0, \ k=1,\ldots, K$ and $\scriptstyle  \sum_{k=1}^K \theta_k=1.$
We can write this probability mass function compactly as:
$$p(z_i|\bm{\theta},K)=\prod_{k=1}^K \theta_k^{I\left( z_i=k\right)},$$
where $I(A)$ is the indicator function $I$ defined as $1$ if condition $A$ is satisfied and $0$ otherwise.
Thus, the distribution of the partition of the $N$ nodes into $K$ clusters conditional on $\bm{\theta}=(\theta_1, \theta_2, \ldots, \theta_K)$ is:
\begin{equation}\label{eq:priorZgivenTheta}
\pi(\bm{z}|\bm{\theta},K)=\prod_{i=1}^N Multi(z_i ; 1,\bm{\theta}) =\prod_{i=1}^N \prod_{k=1}^K \theta_k^{I\left( z_i=k\right)}.
\end{equation}
\paragraph{Prior for $\bm{\theta}$:}
We assume a vague conjugate prior for the vector $\bm{\theta}$ following a Dirichlet distribution of dimension $K$ with vector of concentration parameters $\bm{\gamma}$:	$$\bm{\theta}|K \sim Dir(\bm{\gamma}=(\gamma_1, \gamma_2, \ldots, \gamma_k, \ldots, \gamma_K)).$$
We choose to set all concentration parameters to the same value, thus resulting in a symmetric prior. In particular, we set all $\gamma_k$'s equal to $\gamma_0=1$ yielding a uniform prior. Another possibility would be to set $\gamma_0=\frac{1}{2}$ corresponding to a non informative Jeffrey's prior.
\paragraph{Blocks interaction probabilities:}
Following the stochastic block model framework, the relational pattern of the nodes depends on the block probabilities of each game outcome. The $K\times K\times 3$ array, $\bm{p}$, encoding the block interaction probabilities is therefore of the form:
$$\bm{p}=
\begin{pmatrix}
\underline{p}^{11} & \underline{p}^{12}& \dots & \underline{p}^{1l} & \dots &  \underline{p}^{1K}\\
\underline{p}^{21} & \underline{p}^{22}& \dots & \underline{p}^{2l} & \dots &  \underline{p}^{2K}\\
\vdots & \vdots &\ddots &\vdots& \ddots & \vdots\\
\underline{p}^{k1} & \underline{p}^{k2}& \dots & \underline{p}^{kl} & \dots &  \underline{p}^{kK}\\
\vdots & \vdots &\ddots &\vdots& \ddots & \vdots\\
\underline{p}^{K1} & \underline{p}^{K2}& \dots & \underline{p}^{Kl} & \dots &  \underline{p}^{KK}\\
\end{pmatrix}$$
where, 
\[\underline{p}^{kl}=(p_1^{kl}, p_2^{kl},p_3^{kl})\;\; \mbox{and}\;\; \sum_{\omega=1}^3p_\omega^{kl}=1,\]
for all $k=1,\dots,K$ and $l=1,\dots,K$.
In other words, $p_1^{kl}, p_2^{kl}$ or $p_3^{kl}$ is the probability that a team allocated to block $k$ playing at home against a team allocated to block $l$, wins, draws or loses, respectively. 
Therefore, the random variable denoting the result (win, draw or loss) for any given team in block $k$ playing at home against any team in block $l$, playing away, has the same probability mass function $\underline{p}^{kl}$. 

\paragraph{Distribution of the relational pattern of $\bm{y}$:}
As in \citep{nowicki2001estimation}, we model the distribution of edges between nodes conditionally on the block memberships. Additionally, we model that edges between nodes are identically distributed with parameters given by the interaction matrix $\bm{p}$. In contrast to \citep{come2015model}, where a Bernoulli distribution is chosen, the dyadic relation in our scenario takes categorical values. Here we model the observation $y_{ij}$ conditional on the latent allocations $z_i, z_j$ as a multinomial distribution,
\begin{eqnarray*}
 f(y_{ij}|z_i,z_j,\bm{p},K) &=& \prod_{k=1}^K \prod_{l=1}^K  Multi(y_{ij}; 1,\underline{p}^{kl})^{I{\left(z_i=k\right) I\left( z_j=l\right)}} \\
 &=& \prod_{k=1}^K \prod_{l=1}^K \left\{ \prod_{\omega=1}^3 \left( p_{\omega}^{kl} \right)^{I(y_{ij}=\omega)} \right\}^{I{\left(z_i=k\right) I\left( z_j=l\right)}}
\end{eqnarray*}
for $i,j=1,\dots,N$, $i\neq j$.

Thus, given the latent structure $\bm{z}$, the block interactions array $\bm{p}$ and $K$, we can write the likelihood of the relational pattern $\bm{y}$ as
\begin{eqnarray}
	f(\bm{y}|\bm{z},\bm{p},K) &=& \prod_{i=1}^{N-1} \prod_{\underset{j\neq i}{j=1}}^N f(y_{ij}|z_i, z_j,\bm{p},K) \nonumber \\
	&=& \prod_{i=1}^{N-1} \prod_{\underset{j\neq i}{j=1}}^N \prod_{k=1}^K \prod_{l=1}^K \left\{ \prod_{\omega=1}^3 \left( p_{\omega}^{kl} \right)^{I(y_{ij}=\omega)} \right\}^{I{\left(z_i=k\right) I\left( z_j=l\right)}}.
\end{eqnarray}

\paragraph{Prior for the block interaction probabilities:} 
We assume that the entries of $\bm{p}$ are mutually independent and that each $\underline{p}^{kl}$ follows a conjugate prior from a 3-dimensional Dirichlet distribution:
$$\underline{p}^{kl} \sim Dir(\bm{\beta}=(\beta_1, \beta_2, \beta_3)), \quad \mbox{for}\;\; k=1,\dots,K,\quad \mbox{and}\;\; l=1,\dots,K.$$
We set all the hyperparameters $\beta_1,\beta_2,\beta_3$ to $1$ leading to a uniform distribution.

\paragraph{Prior for $K$:}
We treat the number of blocks or clusters as a random variable and choose a probability mass function for $K$ which is distributed as a zero-truncated Poisson random variable with $\lambda=1$  restricted to $1 \leq k \leq K_{max}$, where $K_{max}$ is a user specified upper limit on the plausible number of blocks. We note that the justification for this prior was developed in \citep{nobile2005} in the context of mixture models and used in several papers including \citep{nobile2007bayesian}, \citep{wyse2012block} and \citep{geng2019}. It consists of a Poisson distribution with rate parameter equal to $1$ conditioned on $K>0$:
$$K\sim Poi(1|K>0),$$
that is,
\begin{equation}
\pi(K|K>0)=\frac{Poi(1)}{1-Poi(K=0)}=\frac{1}{K!(e-1)}.
\end{equation}
Therefore this prior probability mass function is proportional to $\frac{1}{K!}.$ As such, this prior reflects the fact that an SBM with $K$ block contains $K!$ permutations of the block labels, due to the label switching phenomena which is well understood for finite mixture models. This prior therefore assigns equal prior mass to each of these $K!$ possible relabellings. We return to the issue of label switching in Section~\ref{sec:labelswitch}. 
Note that we also explore the case of a discrete uniform prior on $K$ between $1$ and a user specified $K_{max}$. We detail the results of this analysis in application to the English first division/ Premier league in the supplementary material and show that the qualitative results remain largely unchanged compared to using truncated Poisson prior above.

\subsection{Collapsing the stochastic block model}

Following the specification of stochastic block model, we can write out the full joint posterior distribution of all model parameters as
\begin{equation} \label{eq:joint}
\pi(\bm{z},\bm{p},\bm{\theta}, K|\bm{y}) \propto f(\bm{y}| \bm{p},\bm{z},K) \pi(\bm{p}|K) \pi(\bm{z}|\bm{\theta},K) \pi(\bm{\theta}|K) \pi(K).
\end{equation}
Our interest here is primarily the joint posterior distribution of the latent allocation vector and number of blocks, $\pi(\bm{z},K|\bm{y})$. By virtue of the fact that we have chosen a conjugate prior for $\bm{\theta}$ and $\bm{p}$ we can integrate out both of these vectors from the posterior distribution yielding a collapsed posterior distribution,
\begin{align}
\pi(\bm{z},K|\bm{y}) &= \int_{\Theta} \int_{\bm{P}} f(\bm{y}| \bm{p},\bm{z},K) \pi(\bm{p}|K) \pi(\bm{z}|\bm{\theta},K) \pi(\bm{\theta}|K) \pi(K)d\bm{\theta} d\bm{p}\nonumber \\
&=\int_{\bm{P}} f(\bm{y}| \bm{p},\bm{z},K) \pi(\bm{p}|K) d\bm{p} \times \int_{\Theta}  \pi(\bm{z}|\bm{\theta},K) \pi(\bm{\theta}|K) d\bm{\theta} \times \pi(K) \label{eq:collapse1} \\
&= f(\bm{y}|\bm{z},K)\pi(\bm{z}|K) \pi(K). \nonumber
\end{align}
Details of the integration carried out in (\ref{eq:collapse1}) are explained in Appendix A. Following (\ref{eqn:collapseP_final}) and (\ref{eqn:collapse_theta}) the collapsed posterior of the model has the form:
\begin{equation} \label{eq:fulljoint_0}
\pi(\bm{z},K|\bm{y})\propto  \prod_{k=1}^K\prod_{l=1}^K \Gamma(3) \frac{\prod_{\omega=1}^3\Gamma( N_{kl}^{\omega}+1)}{\Gamma(\sum_{\omega=1}^3 \left( N_{kl}^{\omega} +1)\right)} \cdot \prod_{k=1}^K\Gamma(n_k+1)  \frac{\Gamma(K)}{\Gamma(N+K)} \times \frac{1}{K!},
\end{equation}
where we define 
\begin{equation*}
N_{kl}^{\omega} = \sum_{i=1}^{N-1}\sum_{\underset{j\neq i}{j=1}}^N {I(y_{ij}=\omega) I\left(z_i=k\right)I\left(z_j=l\right)}, 
\end{equation*}
for $\omega=1,2,3$ and for $k,=1,\dots,K$ and $l,=1,\dots,K$, therefore allowing for the possibility that $k=l$. Also,  
\begin{equation*}
	n_k=\sum_{i=1}^N{I\left(z_i=k\right)}, \ k=1,\ldots, K. 
\end{equation*}
Therefore $N_{kl}^{\omega}$ counts the number of times that the outcome $\omega$ was observed for all games involving a team allocated to block $k$ playing at home against a team allocated to block $l$. While $n_k$ accounts for the number of nodes/teams allocated to block $k$.

\section{Bayesian estimation of the stochastic block model}
\label{sec:mcmc}

Over time, many inferential strategies have been developed for the stochastic block model and its variants. For example, variational Bayes EM algorithm or integrated likelihood variational Bayes \citep{latouche2012variational}. Moreover, model selection criteria based on the integrated complete likelihood 
\citep{biernacki2000assessing} have been developed for bipartite and binary networks \citep{rastelli2018choosing}. Additionally, \cite{mcdaid2013improved} developed a novel MCMC algorithm for an SBM for a network with binary edges and this is the framework which we develop here for our model. 

The overall objective is to develop an algorithm to sample from the posterior distribution $\pi(\bm{z},K|\bm{y})$~(\ref{eq:fulljoint_0}). In so doing, this will allow us to estimate the posterior distribution of the number of blocks, $\pi(K|\bm{y})$, but also to estimate the posterior distribution $\pi(\bm{z}|K,\bm{y})$, so that we can assign probabilities to the allocation of teams to each of $K$ blocks. We also highlight, similar to \cite{nobile2007bayesian} that although our objective is to estimate the posterior distribution of $K$, we do not require a dimension changing MCMC algorithm such as reversible jump MCMC. This is because the dimension of the model is encoded in the vector $\bm{z}$ which is of fixed dimension $N$. We note that this strategy has also been employed in \cite{wyse2012block} in the context of the latent block model.

We now describe the MCMC algorithm used to sample from (\ref{eq:fulljoint_0}), the posterior distribution $\pi(\bm{z},K|\bm{y})$. The algorithm is based on three move types:
\begin{description}
\item[Move type MK:] Metropolis move to insert or remove an empty cluster. This move changes the current state of $K$ but not the allocation vector $\bm{z}$.
\item[Move type M-GS:] Metropolis-within-Gibbs move that updates all components of the allocation vector $\bm{z}$ but does not change the number of clusters.  
\item[Move type AE:] Metropolis-Hastings move to absorb or eject a cluster. This move affects both $\bm{z}$ and $K$.
\end{description}
We now present an overview of the pseudocode in Algorithm 1  before providing more details on each move type in turn.

 \begin{algorithm}
 	\caption{MCMC algorithm to sample from $\pi(\bm{z},K|\bm{y})$.}
 	\label{alg:MCMC}
 	\begin{algorithmic}[1]
 		\scriptsize
 		\State We begin with an initial state $(\bm{z}^1, K^1)$. 
 		\While {iteration $s<S$}
 		\State With equal probability select a move type MK, M-GS or AE.
 		\If {\emph{MK is selected}}:
 		\If {an \textit{\textbf{insert}} attempt is selected and accepted} 
 		\State $K$ is updated and increased by $1$ and an empty cluster is added
 		\State $\bm{z}^{(s+1)}=\bm{z}^{(s)}$ and $K^{(s+1)}=K^{(s)}+1$
 		\EndIf 
 		\If {a \textit{\textbf{remove}} attempt is selected and accepted} 
	 		\State $K$ is updated and decreased by $1$ and an empty cluster (if there is one) is removed.
	 		\State $\bm{z}^{(s+1)}=\bm{z}^{(s)}$ and $K^{(s+1)}=K^{(s)}-1$
 		\EndIf
 		\State \textit{\textbf{otherwise}}
 		\State The new state is set equal to the current state: $\bm{z}^{(s+1)}=\bm{z}^{(s)}$ and $K^{(s+1)}=K^{(s)}$
 		\EndIf
 		\If {\emph{M-GS is selected}}:
			\State The value of $K$ is unchanged: $K^{(s+1)}=K^{(s)}$
			\State The allocation vector is updated to $\bm{z}^{(s+1)}$
		\EndIf
 		\If {\emph{AE is selected}}:
	 		\State With probability $p_K^e$ or $1-p_K^e$ select an \textit{absorption} attempt or \textit{ejection} attempt, respectively.
	 		\If {an \textit{\textbf{absorption}} attempt is selected and accepted} 
	 		\State $K^{(s+1)}=K^{(s)}-1$ and $\bm{z}$ is updated to $\bm{z}^{(s+1)}$
 		\EndIf
 		\If {an \textit{\textbf{ejection}} attempt is selected and accepted} 
	 		\State $K^{(s+1)}=K^{(s)}+1$ and $\bm{z}$ is updated to $\bm{z}^{(s+1)}$.
 		\EndIf
 		 \textit{\textbf{otherwise}}
	 		\State $\bm{z}^{(s+1)}=\bm{z}^{(s)}$ and $K^{(s+1)}=K^{(s)}$
 		\EndIf
 		\EndWhile
	\end{algorithmic}
\end{algorithm}

\subsection{Move-type MK}

When this move is chosen, the algorithm selects with probability $0.5$ to increase or decrease the number of clusters by inserting or deleting an empty cluster, respectively. In the case of an \textit{insert} proposal, an empty cluster is added and the label with smallest value available is set to this cluster. This move is accepted with probability~(\ref{eqn:insert}).  
When the current number of clusters $K$ is equal to the maximum allowed, $K_{max}$, an \textit{insert} proposal is rejected with probability $1$ and the number of clusters remains at $K=K_{max}$. 
When proposing to \textit{delete} a cluster, the algorithm first checks if some clusters are empty and the cluster with highest label value is selected. This move is accepted with probability~(\ref{eqn:delete}). Otherwise, if $K=1$, the proposal to \textit{delete} a cluster is rejected and the number of clusters remains at $K=1$.
Notice that the move-type MK affects $K$ but leaves the allocation vector $\bm{z}$ unchanged. 
If the proposal to \textit{insert} an empty cluster is accepted, there will be an additional label but no nodes will be assigned to this new cluster. In case of a \textit{remove} proposal, the number of clusters decreases, but again the allocation vector $\bm{z}$ will remain unchanged. Below we provide an algorithmic description of this move type. For more details of the derivation of the acceptance probabilities below, see Appendix B.1.

\newpage
\begin{shaded}
At iteration $s$ with current state $(\bm{z}^{(s)}, K^{(s)})=(\bm{z}, K)$: 
\begin{enumerate}
\item An \textit{insertion} or a \textit{removal} of an empty cluster is proposed with probability $0.5$.
\item If the \textbf{\textit{insertion attempt}} is selected:\\
\begin{itemize}
\item If  $K =K_{max}$, the new state is set equal to the current state: $K^{(s+1)}=K_{max}$.
\item If $K <K_{max}$, we accept $K^{(s+1)}=K+1$ as the new state with acceptance probability: $min \left[ 1, \alpha \right]$, where:
\begin{equation}
  \alpha=\frac{K}{(N+K)(K+1)} 
  \label{eqn:insert}
\end{equation}
\end{itemize}
\item If the \textbf{\textit{delete attempt}} is selected:\\
 \begin{itemize}
	\item If $K=1$, the new state is set equal to the current state: $K^{(s+1)}=1$.
 	\item If $K>1$, we always accept $K^{(s+1)}=K-1$ as the new state since the acceptance probability in this case is $min \left[ 1, \alpha \right]$, where:
\begin{equation}
 \alpha= \frac{K(N+K-1)}{K-1},
 \label{eqn:delete}
\end{equation}
which is always greater than $1$. In other words, we automatically accept the proposal to delete an empty cluster, if there is one.
\end{itemize}
\end{enumerate}

Notice that in all cases the new state for $\bm{z}^{(s+1)}$ will be equal to the current state $\bm{z}^{(s)}$.
\end{shaded}

\subsection{Move-type M-GS}
This move-type involves a standard Metropolis-within-Gibbs update of each element of the allocation vector $\bm{z}$. This sweep consists of updating the allocation of each node sequentially from $z_1$ through to $z_N$. 
Recall that this move-type does not change the current value of $K$. We describe this move type below and provide details of the calculation involved in~(\ref{eqn:m-within-g}) in Appendix B.2 .

\begin{shaded}
At iteration $s$, we carry out a sweep of $\bm{z}^{(s)}$ yielding an updated allocation vector $\bm{z}^{(s+1)}=(z_1^{(s+1)}, z_2^{(s+1)}, \ldots, z_N^{(s+1)})$.
To do this, we update each element $z_i$ of $\bm{z}^{(s)}$ using a Metropolis kernel, for $i=1,\dots,n$, as follows:
\begin{enumerate}
\item Set  $\bm{z}^{(s+1)} = \bm{z}^{(s)}$.
\item For $\{i=1,\dots,n\}$ 
 \begin{enumerate}
 \item Suppose the current state of $z^{(s)}_i=k_0$. Propose a new state $z_i'$ for $z^{(s)}_i$ by sampling a new cluster label uniformly from the set
 $\{1,\dots,K\}\setminus k_0$. 
 \item Denote the proposed new allocation vector as $\bf{z}'$, which is identical to the current state of the allocation vector, $\bm{z}^{(s+1)}$, except for its $i$th element, which is $z_i'$. 
 \item The proposed new allocation vector $\bf{z}'$ is accepted with probability
 \begin{equation}
  \alpha = \min\left( 1, \frac{\pi(\bm{z}',K|\bm{y})}{\pi(\bm{z}^{(s+1)},K|\bm{y})}  \right)
   \label{eqn:m-within-g}
 \end{equation}
 in which case, $z^{(s+1)}_i = z_i'$. Otherwise, $z^{(s+1)}_i = z^{(s)}_i$.
 \end{enumerate}
\end{enumerate}
\end{shaded}

\subsection{Move-type AE}
This move was originally proposed in \cite{nobile2007bayesian} and is designed to allow a change to the number of blocks, $K$ as well as to the allocation vector $\bm{z}$.
It consists of a Metropolis-Hastings pair of absorption/ejection moves. Here we choose to attempt an \textit{ejection} with probability $p_K^e$, depending on the current number of blocks. In particular we set,
\begin{equation*}
p_K^e= \begin{cases}
1, \text{ if } K=1, \\
1/2, \text{ if } 1<K<K_{max}, \\
0 \ , \text{ if } K=K_{max}.
\end{cases} 
\end{equation*}
Otherwise, an \textit{absorption} move is proposed. 

\paragraph{Ejection attempt} A block $j_1$ is randomly selected as the ejecting block among the $K$ available blocks. The ejected block is assigned the label $j_2=K+1$. Each of the nodes in the block $j_1$ are then randomly assigned to a new block $j_2$ or remain in the original block $j_1$.  
The probability for each node to be allocated to the new block $j_2$, that is the probability of a node to be ejected, is constant and indicated with $p_E$, where $p_E\sim U(0,1)$. Note, that instead of specifying an ejection probability $p_E$, we integrate over the $p_E$ in much the same manner as collapsing. The details of this are outlined in Appendix B.3.

\paragraph{Absorption attempt} Here both the absorbing and absorbed block, $j_1$ and $j_2$, respectively, are randomly chosen from the $K$ available labels. All nodes allocated in block $j_2$ are reallocated to block $j_1$.

This move type, as pointed out in \cite{nobile2007bayesian}, leads to improved mixing. 
Reversibility requires that the ejected block is a random draw from the resulting $K+1$ clusters. To achieve this, at the end of the \textit{ejection} move we perform a \textit{swap} between the label $j_2=K+1$ and a randomly selected label from the set of all available ones, $\mathcal{K'}=\{1,2,\ldots, K+1\}$. 
The candidate state $(\bm{z}',K')$ is accepted with probability min$(1,\alpha)$, where $\alpha$ is the product of the joint probability ratio and the proposal ratio as detailed below. 

We detail the proposal ratios and probabilities for the \textit{ejection} and \textit{absorption} moves below. We refer the reader to Appendix B.3 for precise details of the derivation of these quantities.

\begin{shaded}
The ratio of posterior densities for \textit{ejection attempt} is:
\begin{equation}
\frac{ P_{prop}((\bm{z}',K')\rightarrow (\bm{z},K))}{P_{prop}((\bm{z},K)\rightarrow (\bm{z}',K'))}=
\frac{(1-p_K^e)}{p_K^e} \cdot (n_{j_1}+n_{j_2}+1), \label{eq:pinko}
\end{equation}
where,
\vspace{-0.5cm}
\begin{align*}
n_{j_1}&=\sum_{i=1}^N I(z'_i=j_1) \text{ is the number of nodes in the ejecting cluster after reallocation,} \\
n_{j_2}&=\sum_{i=1}^N I(z'_i=j_2) \text{ is the number of nodes in the ejected cluster.}
\end{align*}
To obtain $\alpha$, we multiply (\ref{eq:pinko}) by:
\begin{equation}
\frac{\pi(\bm{z}',K'|\bm{y})}{\pi(\bm{z},K|\bm{y})}=\frac{ \prod_{k=1}^{K+1}\prod_{l=1}^{K+1} \Gamma(3) \frac{\prod_{\omega=1}^3\Gamma( N_{kl}^{' \omega}+1)}{\Gamma(\sum_{\omega=1}^3\left( N_{kl}^{' \omega }+1)\right) } \times \prod_{k=1}^{K+1}\Gamma(n'_k+1)}
{\prod_{k=1}^K\prod_{l=1}^K \Gamma(3) \frac{\prod_{\omega=1}^3\Gamma( N_{kl}^{\omega}+1)}{\Gamma(\sum_{\omega=1}^3\left( N_{kl}^\omega +1) \right) } \times \prod_{k=1}^K\Gamma(n_k+1)  } \cdot \frac{K}{(N+K)(K+1)}
\end{equation}
and accept this move with probability $\min(1,\alpha)$.
\end{shaded}

\begin{shaded}
For the \textit{absorption attempt} the proposal ratio takes the form:
\begin{equation}
\frac{ P_{prop}((\bm{z}',K')\rightarrow (\bm{z},K))}{P_{prop}((\bm{z},K)\rightarrow (\bm{z}',K'))}=
\frac{p_K^e}{(1-p_K^e)} \cdot \frac{1}{(n_{j_1}+n_{j_2}+1)}.
\label{absorb_1}
\end{equation}
The corresponding posterior density ratio appears as:
\begin{equation}
\frac{\pi(\bm{z}',K'|\bm{y})}{\pi(\bm{z},K|\bm{y})}=\frac{ \prod_{k=1}^{K-1}\prod_{l=1}^{K-1} \Gamma(3) \frac{\prod_{\omega=1}^3\Gamma( N_{kl}^{' \omega}+1)}{\Gamma(\sum_{\omega=1}^3\left( N_{kl}^{' \omega}+1) \right) } \times \prod_{k=1}^{K-1}\Gamma(n'_k+1) }
{\prod_{k=1}^K \prod_{l=1}^K \Gamma(3) \frac{\prod_{\omega=1}^3\Gamma( N_{kl}^\omega+1)}{\Gamma(\sum_{\omega=1}^3\left( N_{kl}^\omega +1) \right) } \times \prod_{k=1}^K \Gamma(n_k+1) } \cdot \frac{K(N+K-1)}{K-1}.
\label{absorb_2}
\end{equation}
Once again, $\alpha$ is then the product of (\ref{absorb_1}) and (\ref{absorb_2}) and the move is accepted with probability $\min(1,\alpha)$.
\end{shaded}

\subsection{Label switching phenomenon and correction}
\label{sec:labelswitch}
In the framework outlined above, the discrete labels in the allocation vector $\bm{z}$ are not identifiable by the model. This is because the likelihood is invariant to permutations of the labels of $\bm{z}$, leading to the well-known label switching phenomenon. This scenario arises naturally when using MCMC methods and, as a result, it is necessary to employ an algorithm to correct for this. 
In particular we use the online relabelling algorithm proposed in \cite{nobile2007bayesian}  which relies on the algorithm of \cite{carpaneto1980algorithm} to correct the output of a sample from the posterior distribution of $\bm{z}$ from the MCMC algorithm described above. The relabeling algorithm which we use first orders this sample by the number of non-empty blocks in increasing order to yield an ordered sample $\bm{Z} := \{\bm{z}^{(1)}, \bm{z}^{(2)},\ldots,\bm{z}^{(S)}\}$. The algorithm then iterates over $\bm{Z}$ comparing the current allocation vector to all previously relabeled states. The current state is then relabeled by permuting its labels so that a distance to all the preceding relabeled states is minimised. Here we define a distance between any two allocation vectors, $\bm{z}$ and $\bm{z}^{\prime}$ as
\begin{equation} 
 D(\bm{z},\bm{z}^{\prime})=\sum_{i=1}^N I(z_i \neq z_i^{\prime}), 
 \label{eqn:distance}
\end{equation}
where $I$ is the usual indicator function used previously, so that this distance records the number of locations where the entries of $\bm{z}$ and $\bm{z}^{\prime}$ differ. Further, let $\sigma$ denote a permutation of the integers $1,\dots,K$, that is, a bijection from the set $\{1,\dots,K\}$ onto itself, and let $\sigma\bm{z}$ denote the relabeled allocation vector by applying the permutation to the labels of the vector $\bm{z}$. This algorithm was implemented using the \texttt{R} package \texttt{collpcm} and can be briefly described as follows:
\begin{shaded}
\begin{enumerate}
	\item The sample of $S$ allocation vectors from the MCMC algorithm are ordered by the number of non-empty blocks and in decreasing order with respect to the number of labels used yielding an sequence of allocation vectors  $\{\bm{z}^{(1)}, \bm{z}^{(2)},\ldots,\bm{z}^{(S)}\}$.
	\item Set $\sigma^{(1)}$ to be the identity permutation.
	\item For $\{s=2,\dots,S\}$ \\ 
	 Find the permutation $\sigma^{(s)}$ which minimises the sum of distances from the permuted allocation vector $\bm{z}^{(s)}$ to all preceding permuted allocation vectors, 
	 \[
	  \sigma^{(s)} = \arg\min_{\sigma} \sum_{t=1}^{s-1} D(\sigma \bm{z}^{(s)}, \sigma^{(t)}\bm{z}^{(t)}),
	 \]
    where $D(\sigma \bm{z}^{(s)}, \sigma^{(t)}\bm{z}^{(t)})$ is the distance function (\ref{eqn:distance}).
	\item The relabeled sequence of allocation vectors 
	\[ 
	  \{\sigma^{(1)}\bm{z}^{(1)},\dots,\sigma^{(S)}\bm{z}^{(S)} \}
	\]
	is returned. 
\end{enumerate}
\end{shaded}

\section{Analysis of over $40$ seasons of the English First Division/ Premier League}
\label{sec:results}

In this section, we present a detailed analysis for the 2021/22 Premier League season, highlighting some of the important aspects of the output from the model. 
In particular, for this season, the SBM partitioned the teams into two blocks where one can visually see structure within and between each block -- namely that teams within each block are quite competitive while when a team plays another from a different block, the result usually favours the team from the strongest block.
We then extend our analysis to each season over the past $44$ years from $1978/79$ to $2021/22$ encompassing the end of the old English First Division through to the inception of the Premier League. Analysing each season over this length of time will allow us to provide insights into changes in competitive balance.  
For each season, the MCMC algorithm described in Section~\ref{sec:mcmc} was run for $200,000$ iterations, discarding the first $50,000$ as burn-in iterations and took approximately $3$ minutes to complete. This was followed by the label switching correction algorithm which took only a few seconds to run.   
The MCMC algorithm was coded in \texttt{R} and a link to this code and the data used in this paper can be found here \url{https://github.com/francescabasini/Football_SBM}.

\subsection{Analysis of the 2021/22 Premier League season} 
\label{sec:18/19} 
 
\begin{figure}[H]
\centering
\includegraphics[scale=0.2]{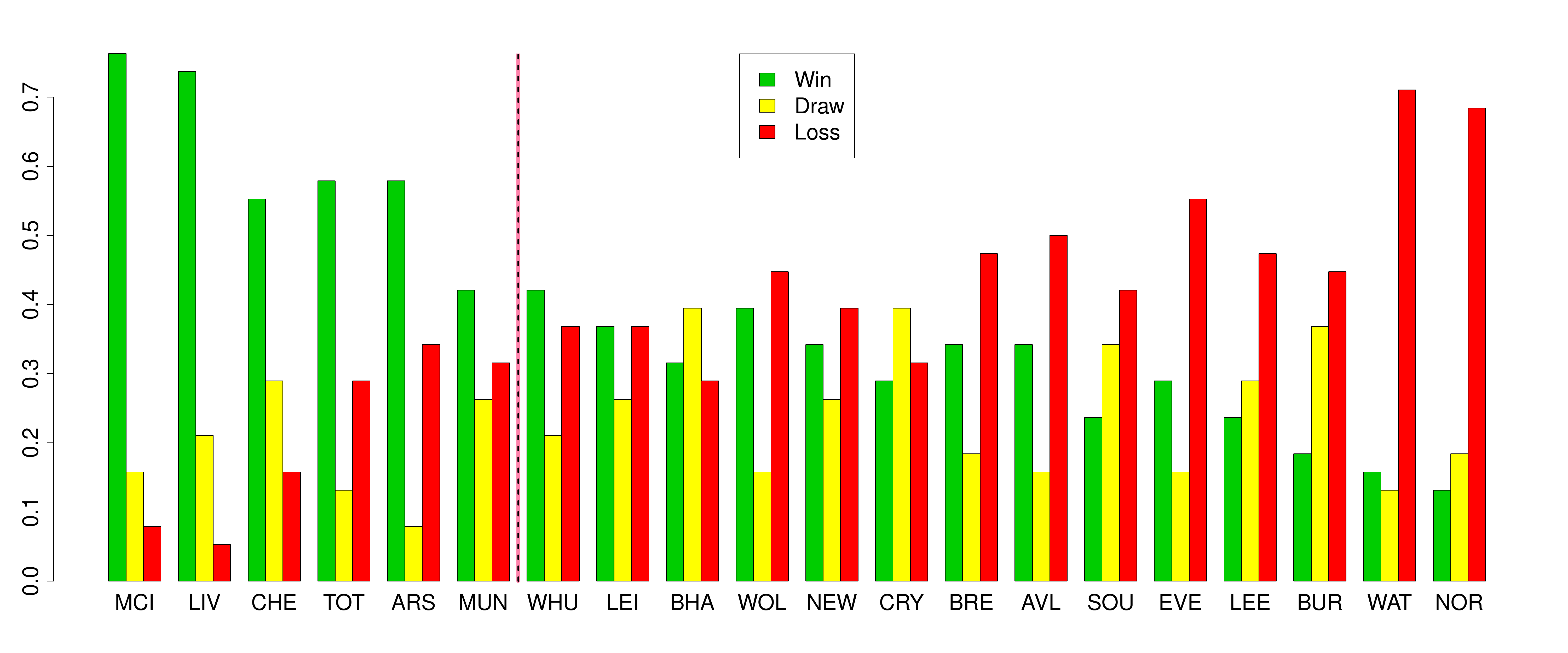}
\caption{Barplot of the percentage of overall outcomes (win, draw, loss) for each team for Season 2021/22. Teams are listed in decreasing order from left to right according to their position in the final league table. The dotted line corresponds to where teams are separated into their most likely block. \\
} 
\label{fig:Barplot_overall}
\end{figure}

Here we present a detailed analysis of the $2021/22$ season as this provides us with the opportunity to discuss some of the salient features of our model. To begin we present in Figure~\ref{fig:Barplot_overall} a summary of the outcome for each team in terms of the percentage of wins, draws and losses over the course of the season. As expected, this illustrates that teams that finish higher in the league tend to have more wins. While the opposite happens for teams which are placed lower in the league. It also illustrates the disparity in terms of percentage of wins (and losses) between teams with a high final position in the league compared to those whose final position was towards the bottom of the league.

\begin{figure}
	\begin{tabular}{cc}
		\centering
		\includegraphics[scale=0.33]{Heatmap_Season_2122_final.pdf} &
		\includegraphics[scale=0.33]{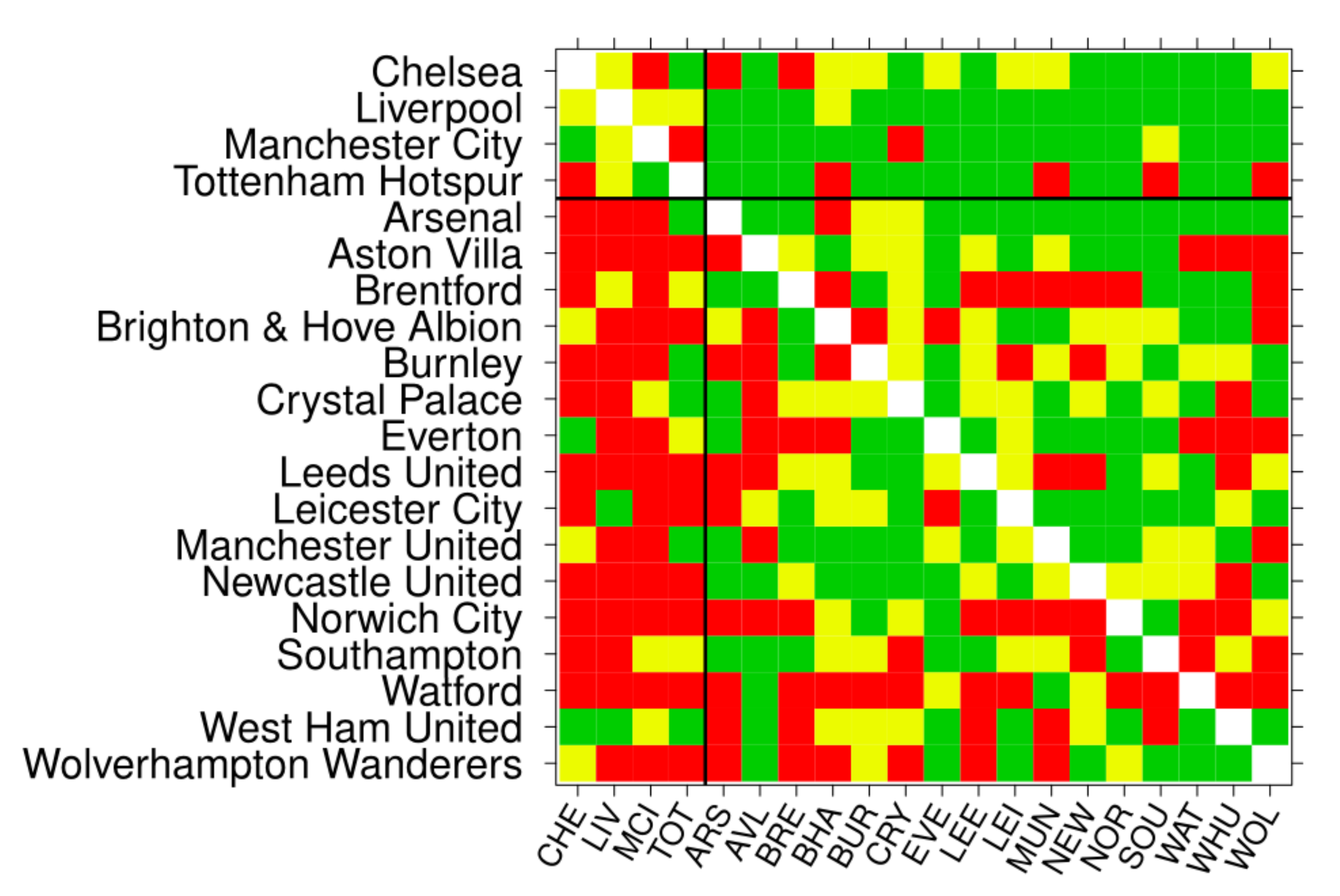} \\
		(a) & (b) \\
	\end{tabular}
	\caption{Match grid for the 2021/22 season. Each entry of this grid corresponds to a match where a home team (row entry) plays against an away team (column entry). The outcome of each match is represented using green, yellow and red colours corresponding to a win, draw and loss, respectively, for the home team. (a) Teams listed in alphabetical order. (b) Teams listed by most likely block membership, a posteriori. The solid horizontal and vertical lines (which also coincides with final league position) separates each team in their most likely block.
	}
	\label{fig:season2122_block_coloured1}%
\end{figure}

As before, the data for this season is presented in Figure~\ref{fig:season2122_block_coloured1} (a) in the form of a results grid.

The MCMC algorithm~\ref{alg:MCMC} was implemented using each of the three moves types, MK, GS and AE following the description in Section~\ref{sec:mcmc}.
The posterior probability of $K=1$, $K=2$ and $K=3$ was estimated to be $0.0$, $0.97$ and $0.03$, respectively. This indicates that there is overwhelmingly most support, a posteriori, for a model with two blocks. 
While the posterior distribution of the allocation variable $\bm{z}$ conditional on $K=2$ indicates that the strongest block is comprised of Chelsea, Liverpool,  Manchester City, Tottenham. 
We also also illustrate this in Table~\ref{table:Grid2122} by presenting the estimated posterior probability that each team belongs to the strongest block, conditional on $K=2$, together with the overall points total for each team at the end of the season.

Note that further output of the model for this season is presented in Appendix C of the supplementary material.
We also note that permuting the rows and columns of the match grid matrix in Figure~\ref{fig:season2122_block_coloured1} (b) by final league position reveals the block structure of the results. Here the solid vertical and horizontal lines separate the two blocks of teams. We make the following remarks to illustrate some important features of the estimated block structure in Figure~\ref{fig:season2122_block_coloured1} (b).
The top right quadrant corresponds to when a team in the strongest block plays at home against a team in the weakest block. Here the colours are predominately green, indicating that the home team generally won.
While the bottom left quadrant corresponds to the opposite situation, when a team in the weakest block plays at home against a team in the strongest block. Here the colours are predominately red, indicating the away team (in the strongest block) generally won.
Finally, note that the top left and bottom right quadrants, corresponds to the within-block matches for the strong block and weak blocks, respectively.
 Each block contains a reasonably balanced mix of all three colours, indicating that these teams were generally quite balanced.

\begin{table}
	\begin{tabular}{rr r}
		\toprule
		& Points & $P(\mbox{top block}|\bm{y},K=2)$ \\ 
		\midrule
		Manchester City  & 93 & 1.00 \\ 
		\rowcolor{black!10}Liverpool  & 92 & 0.99\\ 
		Chelsea  & 74 & 0.88\\ 
		\rowcolor{black!10}Tottenham Hotspur  & 71 & 0.74\\ 
		\toprule
		Arsenal  & 69 & 0.38\\ 
		\rowcolor{black!10}Manchester United  & 58 & 0.02 \\ 
		West Ham United  & 56 & 0.01\\ 
		\rowcolor{black!10}Leicester City  & 52 & 0.00\\ 
		Brighton \& Hove Albion  & 51 & 0.01\\ 
		\rowcolor{black!10}Wolverhampton Wanderers  & 51 & 0.01\\ 
		Newcastle United  & 49 & 0.00 \\ 
		\rowcolor{black!10}Crystal Palace  & 48 & 0.00\\ 
		Brentford  & 46 & 0.00 \\ 
		\rowcolor{black!10}Aston Villa  & 45 & 0.00 \\ 
		Southampton  & 40 & 0.00\\ 
		\rowcolor{black!10}Everton  & 39 & 0.00\\ 
		Leeds United  & 38 & 0.00\\ 
		\rowcolor{black!10}Burnley  & 35 &0.00 \\ 
		Watford  & 23 & 0.00\\ 
		\rowcolor{black!10}Norwich City  & 22 & 0.00\\ 
		\bottomrule
	\end{tabular}
	
	\caption{Final league table for the Premier League Season 2021/22 showing the number of points for each team. For comparison, the estimated posterior probability of allocation of each team to the top block is also presented. The solid horizontal line separates the $4$ teams most likely to be in the top block, a posteriori. \\
	}
	\label{table:Grid2122}%
	
\end{table}

\subsubsection{Analysing the posterior distribution of $\underline{p}^{kk}$ to assess evidence of a home effect}

It is possible to use the output of the model to infer the posterior distribution of $\bm{p}$ conditional on the latent allocation $\bm{z}$ and $K$. It particular, this allows us to address the question of whether there is an advantage to playing at home, by analysing the posterior distribution of the within block probability vector, $\underline{p}^{kk}$, for $k=1,\dots,K$. It is useful to do so, as teams within a block should be balanced and so any effect of playing at home should be more apparent.
It turns out that this posterior distribution of $\bm{p}$ is proportional to the likelihood (given by equation (4) in the text),
\[
 \pi(\bm{p}|\bm{y},\bm{z},K) \propto \prod_{i=1}^{N-1} \prod_{\underset{j\neq i}{j=1}}^N \prod_{k=1}^K \prod_{l=1}^K \left\{ \prod_{\omega=1}^3 \left( p_{\omega}^{kl} \right)^{I(y_{ij}=\omega)} \right\}^{I{\left(z_i=k\right) I\left( z_j=l\right)}}.
\]
In turn, this allows us to sample the within block multinomial probabilities, $\underline{p}^{kk}$ by conditioning on teams only allocated to block $k$. In the context of the $2021/22$ season, it turns out that since block $1$ consists of only $4$ teams, there are only $12$ games involving teams from this block of which there were $3$ home wins, $6$ draws and $3$ home losses. In turn, this is reflected in the inferred posterior distribution of $\underline{p}^{11}$ presented in Table~\ref{table:within_block} which shows that the posterior density for $p^{11}_1$ is identical to $p_3^{11}$. By contrast, block $2$ contains many more teams. From Table~\ref{table:within_block} we see that the posterior density of $p^{22}_1$ has a $95\%$ credible interval of $[0.36, 0.49]$ which is much greater than the corresponding $95\%$ credible interval for $p^{22}_3$ which is estimated to be $[0.25, 0.37]$. This indicates the presence of a home effects within this block of teams.

\begin{table}[H]
\begin{tabular}{rrrrr}
\toprule
     & mean & sd &  $2.5\%$ & $97.5\%$ \\ 
     \midrule
$p_1^{11}$ & 0.26    &   0.11 & 0.08 &   0.50 \\
\rowcolor{black!10}$p_2^{11}$ & 0.47 & 0.12 & 0.24 & 0.71    \\
$p_3^{11}$ & 0.27     &  0.11 & 0.09 &  0.50  \\
\midrule
$p_1^{22}$ & 0.42 & 0.03 & 0.36 &  0.49\\
\rowcolor{black!10}$p_2^{22}$ & 0.27 & 0.03 & 0.22 &  0.33    \\
$p_3^{22}$ & 0.31    &   0.03 & 0.25 &  0.37  \\
\midrule
$p_1^{12}$ & 0.75  &  0.05 & 0.64 &  0.84\\
\rowcolor{black!10}$p_2^{12}$ & 0.13  &  0.04 & 0.06 & 0.22    \\
$p_3^{12}$ & 0.12 & 0.04 & 0.05 &  0.21  \\
\midrule
$p_1^{21}$ & 0.15 & 0.04 & 0.08 & 0.24 \\
\rowcolor{black!10}$p_2^{21}$ & 0.16   & 0.05 & 0.08 &  0.26 \\
$p_3^{21}$ & 0.69 & 0.06 & 0.57 &  0.79  \\
\bottomrule
\end{tabular}
\caption{Posterior summaries (mean, standard deviation, $2.5$th and $97.5$th percentiles) of the multinomial probability vectors, $\underline{p}^{11}$, $\underline{p}^{22}$,$\underline{p}^{12}$ and $\underline{p}^{21}$, respectively. 
}
\label{table:within_block}
\end{table}

Additionally, as one might expect, the posterior density for $\underline{p}^{12}$ indicates that the probability of a win, when a home team in block $1$ plays a team in block $2$, is very high, with a $95\%$ credible interval $[0.64, 0.84]$. While the opposite is the case when a team in block $2$ plays at home against a team in block $1$. Here the $95\%$ credible interval for $p^{21}_1$ is $[0.08, 0.24]$. 

\subsection{Analysis of four decades of the English first division/ Premier League}

Here we extend our analysis to each season over the past $44$ years from $1978/79$ to $2021/22$. This period of time encompasses the end of the old English First Division and the inception of the Premier League in $1992/93$ and allows an examination of any changes in the competitive balance of the league. We begin by providing a summary of the output of our model for each season exploring the posterior probability of $K$ for each season. A summary of the output of our analysis on a season-by-season basis is presented in Table~\ref{table:allResults} where we displayed the estimated posterior probability for a one block model through to a four block model. 
We then extend this analysis in Section~\ref{sec:top_block} to provide more detail on the block structure for each season by analysing the marginal posterior probability of allocation of teams to the strongest block. This allows us to explore how the number of teams in the strongest block has changed over time. In particular, in Section~\ref{sec:topsix} we explore which teams have persisted in the strongest block over the past two decades of our study.
This then facilitates an assessment of whether there is evidence to suggest that a \textit{big-six} group of teams emerged over this time period. 

\begin{table}
		\centering
		\begin{tabular}{cc}

			\begin{tabular}{c|cccc}
				\hline
				& \multicolumn{2}{c}{Number of clusters}\\
				Season & 1 & 2 & 3 & 4 \\ 
				\hline
				78/79 & 1.17 & \cellcolor{powderblue(web)} 96.74 & 2.08 & 0.01 \\ 
                79/80 & \cellcolor{navajowhite2}97.57 & 2.39 & 0.04 & 0.00 \\ 
                80/81 & 30.55 & \cellcolor{powderblue(web)} 69.10 & 0.35 & 0.00 \\ 
                81/82 & \cellcolor{navajowhite2}97.25 & 2.72 & 0.02 & 0.00 \\ 
                82/83 & \cellcolor{navajowhite2}99.80 & 0.20 & 0.00 & 0.00 \\ 
                83/84 & \cellcolor{navajowhite2}99.15 & 0.85 & 0.00 & 0.00 \\ 
                84/85 & 42.34 & \cellcolor{powderblue(web)} 57.22 & 0.44 & 0.00 \\ 
                85/86 & 0.00 & \cellcolor{powderblue(web)} 99.81 & 0.19 & 0.00 \\ 
                86/87 & \cellcolor{navajowhite2}99.49 & 0.51 & 0.00 & 0.00 \\ 
                87/88 & 12.41 & \cellcolor{powderblue(web)} 87.26 & 0.33 & 0.00 \\ 
                88/89 & \cellcolor{navajowhite2}99.20 & 0.79 & 0.01 & 0.00 \\ 
                89/90 & \cellcolor{navajowhite2}98.66 & 1.32 & 0.02 & 0.00 \\ 
                90/91 & 49.31 & \cellcolor{powderblue(web)}50.07 & 0.61 & 0.00 \\ 
                91/92 & \cellcolor{navajowhite2}94.10 & 5.89 & 0.01 & 0.00 \\ 
                92/93 & \cellcolor{navajowhite2}98.85 & 1.15 & 0.00 & 0.00 \\ 
                93/94 & 27.98 & \cellcolor{powderblue(web)}71.08 & 0.94 & 0.00 \\ 
                94/95 & 22.19 & \cellcolor{powderblue(web)}74.59 & 3.21 & 0.01 \\ 
                95/96 & 48.05 & \cellcolor{powderblue(web)}51.68 & 0.27 & 0.00 \\ 
                96/97 & \cellcolor{navajowhite2}99.63 & 0.37 & 0.00 & 0.00 \\ 
                97/98 & \cellcolor{navajowhite2}98.14 & 1.85 & 0.01 & 0.00 \\ 
                98/99 & 0.07 & \cellcolor{powderblue(web)}99.78 & 0.16 & 0.00 \\ 
                99/00 & \cellcolor{navajowhite2}64.34 & 35.27 & 0.39 & 0.00 \\ 
				\hline
			\end{tabular} 
			& \hspace*{0.6cm} 
			\begin{tabular}{c|cccc}
				\hline
				& \multicolumn{2}{c}{Number of clusters}\\
				Season & 1 & 2 & 3 & 4 \\ 
				\hline
				00/01 & \cellcolor{navajowhite2}88.26 & 11.06 & 0.68 & 0.00 \\
				01/02 & 0.13 & \cellcolor{powderblue(web)}99.37 & 0.49 & 0.01 \\ 
                02/03 & \cellcolor{navajowhite2}59.29 & 40.19 & 0.52 & 0.01 \\ 
                03/04 & 6.45 & \cellcolor{powderblue(web)}88.39 & 5.12 & 0.04 \\ 
                04/05 & 0.00 & \cellcolor{powderblue(web)}99.79 & 0.21 & 0.00 \\ 
                05/06 & 0.15 & \cellcolor{powderblue(web)}97.08 & 2.73 & 0.04 \\ 
                06/07 & 5.45 & \cellcolor{powderblue(web)}92.37 & 2.17 & 0.01 \\ 
                07/08 & 0.00 & \cellcolor{powderblue(web)}93.81 & 5.92 & 0.27 \\ 
                08/09 & 0.00 & \cellcolor{powderblue(web)}99.24 & 0.76 & 0.00 \\ 
                09/10 & 0.00 & \cellcolor{powderblue(web)}95.30 & 4.67 & 0.03 \\ 
                10/11 & \cellcolor{navajowhite2}79.87 & 20.09 & 0.05 & 0.00 \\ 
                11/12 & 1.68 & \cellcolor{powderblue(web)}96.91 & 1.40 & 0.02 \\ 
                12/13 & 0.00 & \cellcolor{powderblue(web)}99.64 & 0.36 & 0.00 \\ 
                13/14 & 0.00 & \cellcolor{powderblue(web)}98.96 & 1.04 & 0.00 \\ 
                14/15 & 7.30 & \cellcolor{powderblue(web)}88.70 & 3.98 & 0.03 \\ 
                15/16 & \cellcolor{navajowhite2}75.37 & 24.38 & 0.25 & 0.00 \\ 
                16/17 & 0.00 & \cellcolor{powderblue(web)}98.99 & 1.01 & 0.00 \\ 
                17/18 & 0.00 & \cellcolor{powderblue(web)}97.95 & 2.03 & 0.02 \\ 
                18/19 & 0.00 & \cellcolor{powderblue(web)}97.94 & 2.04 & 0.02 \\ 
                19/20 & 2.10 & \cellcolor{powderblue(web)}96.47 & 1.41 & 0.01 \\ 
                20/21 & 15.86 & \cellcolor{powderblue(web)}81.85 & 2.25 & 0.04 \\ 
                21/22 & 0.04 & \cellcolor{powderblue(web)}96.58 & 3.35 & 0.03 \\ 
				\hline
			\end{tabular}
		\end{tabular}
		\caption{Posterior probability of $K$, expressed as a percentage, over the last $44$ seasons. The block model with highest posterior probability is coloured accordingly.}
		\label{table:allResults}
\end{table}

Overall, Table~\ref{table:allResults} indicates there was most support each season, a posteriori, for either a one block or a two block model, however the number of seasons where a two block model has most support, a posteriori, increases considerably over the past two decades. In particular, over the first half of this study period there is no strong support for either a one or a two block model. In fact for some seasons there is broadly equal posterior support for either model, for example, seasons $1984/85$, $1990/91$, $1995/96$. In fact, of note is season $1984/85$ where the second block consisted of a single team, Stoke City, who were bottom of the league in that season recording only $3$ wins from a possible $42$ matches.
This situation changes considerably when we analyse the final two decades of this study. The right hand side of Table~\ref{table:allResults} indicates that there is typically most support from a two block model, but further that the posterior probability for a two block model is over $0.8$ for almost every season since $2003/04$, providing strong evidence that the league has become more competitively imbalanced since then. There are a few exceptions to this, most notably, season $2015/16$ when Leicester City famously won their first league title. However, in general, this analysis suggests a structural change in the nature of the competitiveness of the league since the turn of the millennium. 

Additionally, there was quite limited support for a three block model, for example, from the beginning of the study until $2002/03$ almost all seasons presented very little, if any, posterior support for this model, with the exception of season $1994/95$, for which a three block model was attributed $0.03$ probability. Conversely, from $2003/04$ to the end of the study, there were some seasons which gave some modest support to a three block model. For example, in seasons $2003/04$, $2007/08$, $2009/10$, there was approximately $0.05$ posterior probability for this model. Again, this is consistent with the idea that the league was more imbalanced over the past two decades. 

\subsubsection{Marginal posterior allocation of a team to the strongest block}
\label{sec:top_block}

\begin{figure}
	\centering
	\includegraphics[scale=0.33]{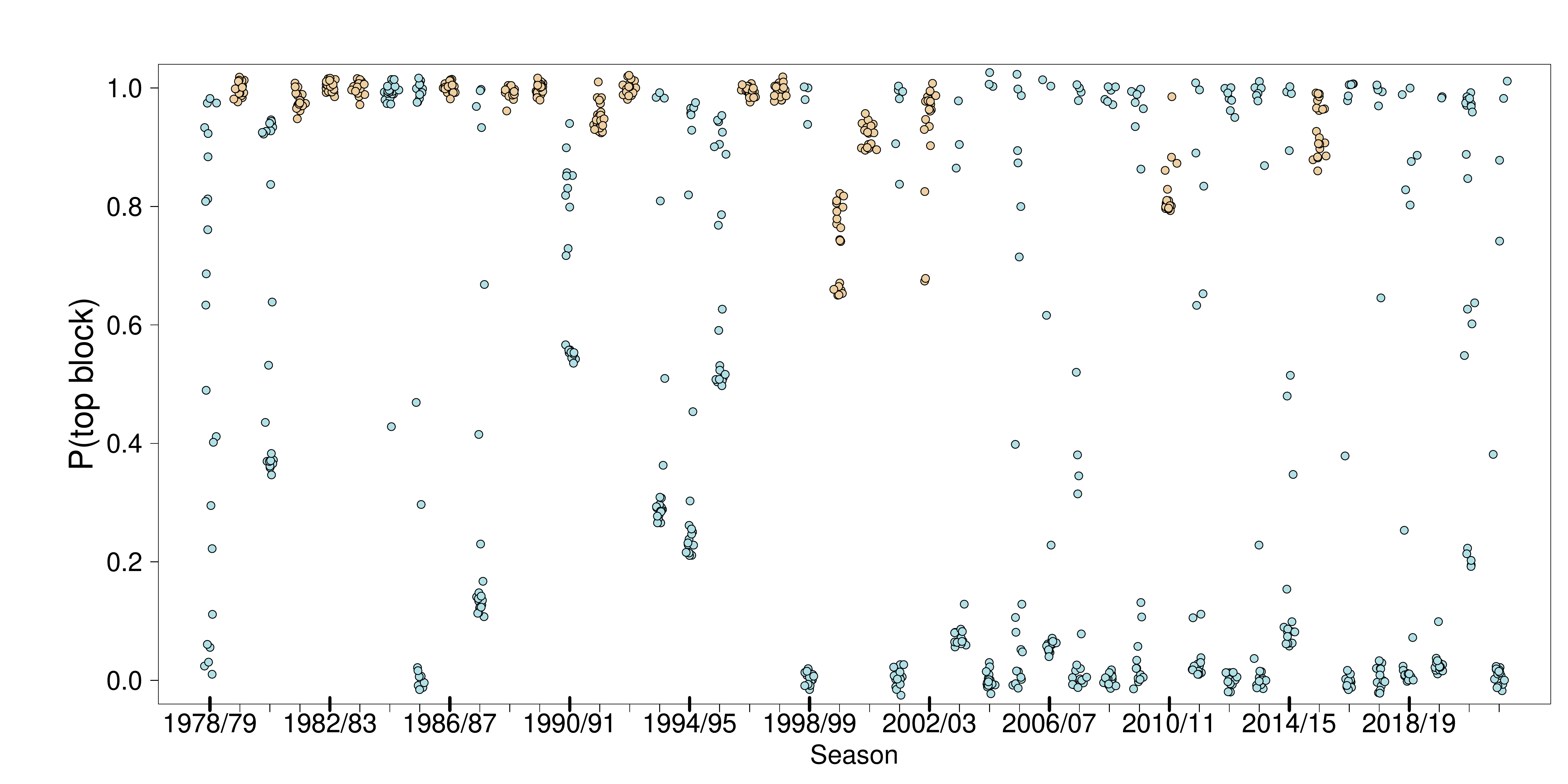}
	\caption{The posterior allocation probability of belonging to the strongest group of teams over the $44$ seasons under study. For each season the colour indicates whether the league was partitioned into a single cluster (sand colour) or two (light blue). Each point represents the estimated posterior probability of allocation for a team. Note that within each season the points have been jittered.}
	\label{fig:beanplot}
\end{figure}

It is informative to estimate the marginal probability that a team belongs to the strongest block by integrating over the uncertainty in the number of blocks. This is particularly important for seasons where there is broadly equal support for a one or a two block model. For each value of $k$, post label switching we associate the block label $1$ with the strongest block of teams. We estimate the marginal posterior probability that team $i$ is in the strongest block averaging over all possible number of blocks as follows,
\begin{eqnarray}
 \pi(z_i=1|\bm{y}) &=& \sum_{k=1}^{K_{max}}\pi(z_i=1, K=k|\bm{y}) \nonumber \\
 &=& \sum_{k=1}^{K_{max}} \pi(z_i=1|\bm{y},K=k) \pi(K=k|\bm{y}). 
\label{eqn:marg_top_block}
\end{eqnarray}
In Figure~\ref{fig:beanplot} we display the estimated marginal posterior allocation of each team to the strongest block following~(\ref{eqn:marg_top_block}) for each season. Moreover, we again use the convention to colour each season according to the block model with most support (sand and blue corresponding to seasons in which a one block model or a two block model, respectively, has most support). 
Notice in Figure~\ref{fig:beanplot} that for seasons from $1978/79$ to around $2002/03$ the spread of posterior probability of allocation to the top block for each team is quite different to that of the seasons which follow. In particular, from $2003/04$ to $2021/22$ we see that there are typically a relatively large group of teams (in a two block season) which have very small posterior probability of being allocated to the strongest block, indicated in the bottom right hand corner of Figure~\ref{fig:beanplot}. In other words, there is a strong separation of teams in terms of posterior allocation to the strongest block, compared to the first half of seasons in this study.  
In addition, we make the important remark that the number of teams allocated the strongest block for each of the seasons from $2003/04$ to $2021/22$, that is to say, those with high posterior probability of allocation to the strongest block, typically consists of a small number of teams. 
This analysis leads us in Section~\ref{sec:topsix} to explore the composition of teams in the strongest block over the past two decades.

\begin{figure}
	\centering
\includegraphics[scale=0.33]{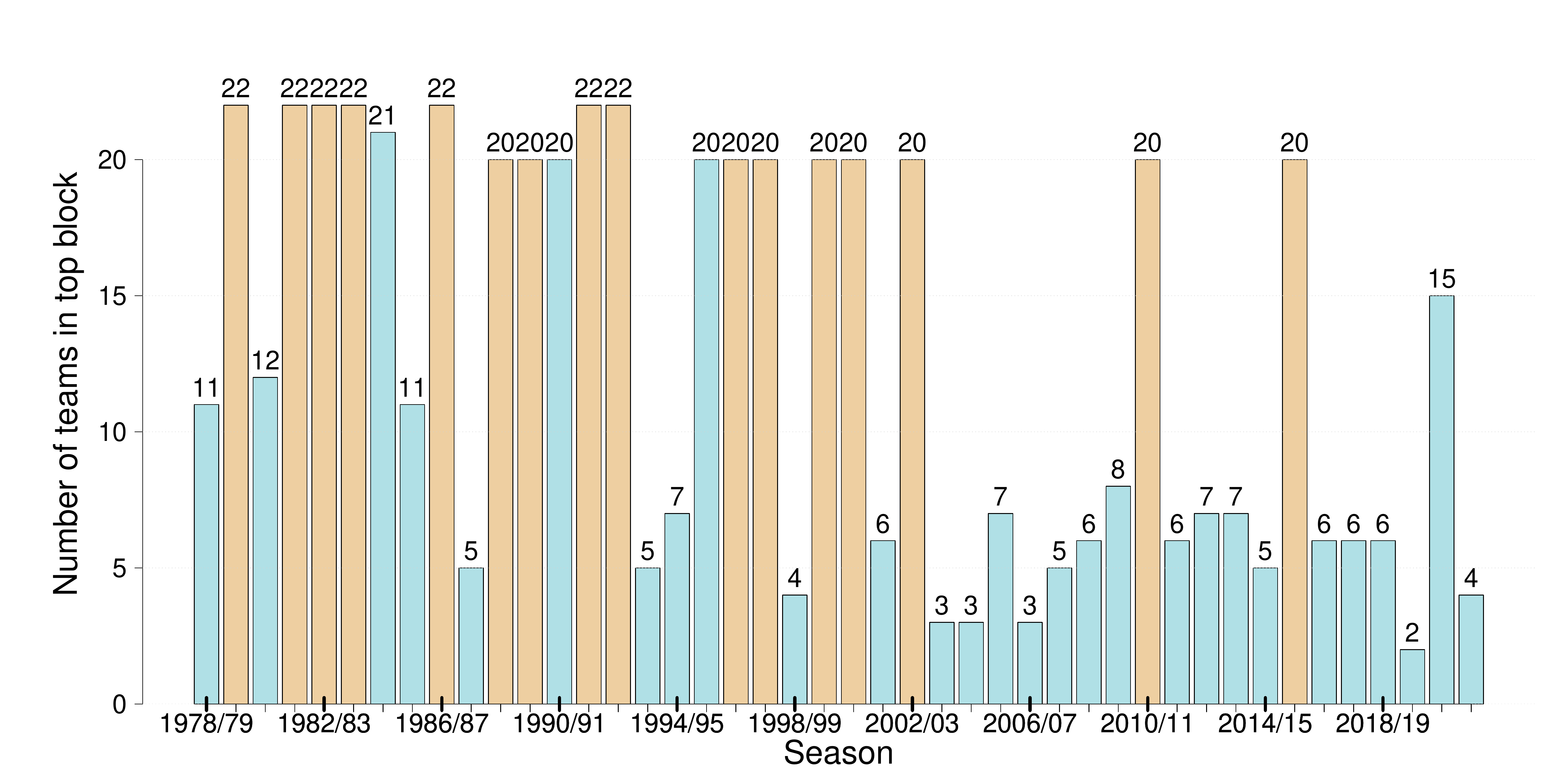}
	\caption{Barplot displaying the posterior estimate of the number of teams allocated to the strongest block each season. Single and two block seasons are coloured sand and blue, respectively. This illustrates that the size of the strongest block has generally decreased during the second half of the study period.
	}
	\label{fig:topblocksize}
\end{figure}

Following from the analysis presented above we can explore the number of teams estimated to belong to the strongest block, by again integrating over the uncertainty in $k$. To do this, we simply examine~(\ref{eqn:marg_top_block}) for each team and investigate if this posterior probability is greater than $0.5$. If so, we assign this team to the strongest block of teams. For each season, this allows one to estimate the size of the strongest block of teams for each season in turn. This number serves as a useful alternative index of competitive balance for each season. This information is presented in Figure~\ref{fig:topblocksize}. We highlight for certain seasons that although a two block model has most posterior support, the estimated number of teams in the strongest block can be close to or equal to the total number of teams in the league that season. This is the case for seasons $1990/91$ and $1995/96$ where there was broadly equal posterior support for a one and a two block model, integrating over the uncertainty in the number of blocks indicates that all $20$ are estimated to be in the strongest block suggesting quite a balanced league for these seasons.

An overall analysis of Figure~\ref{fig:topblocksize} indicates that the number of teams in the strongest block is quite large for the majority of seasons in the first half of the study period until around $2003/04$. From $1978/79$ to $2003/04$ the estimated size of the top block contained more than half the total number of teams in the league each season, ranging from $11$ to $22$ teams for $20$ out of $25$ season. 
This is in stark contrast with the seasons which followed, where the size of the top block ranged from $2$ to $8$ teams for $15$ out of $18$ season. A notable exception is season $2020/21$, where the size of top block was $15$ teams. In this season COVID-19 had an impact on the league, particularly through the absence of crowds for the majority of games in this season.

\begin{figure}
		\centering-
		\floatbox{figure}{
			\begin{subfloatrow}[2]
				\ffigbox[\FBwidth]{
					\includegraphics[scale=0.15]{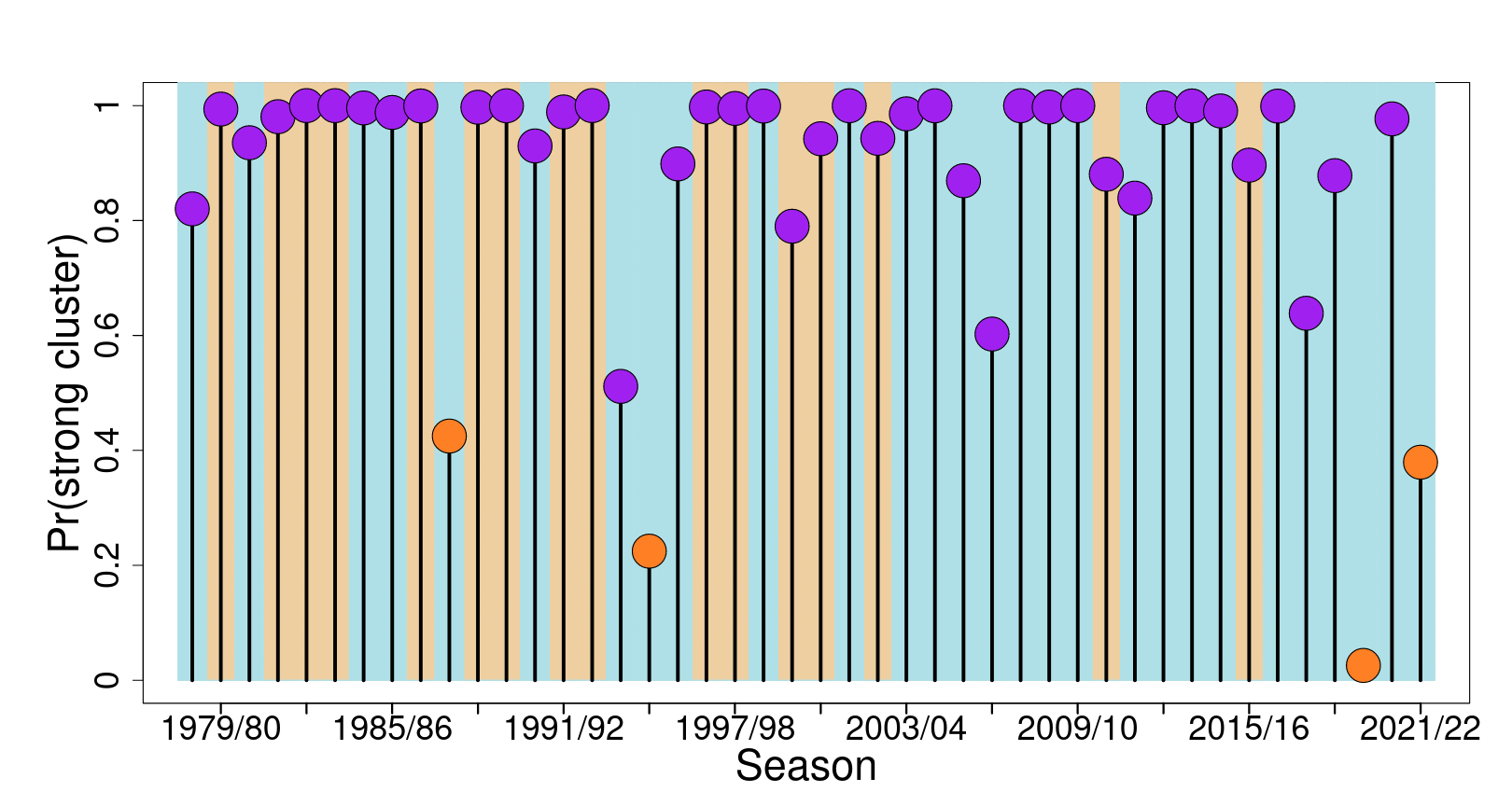}}{
					\caption{Arsenal.}}
				\ffigbox[\FBwidth]{
					\includegraphics[scale=0.15]{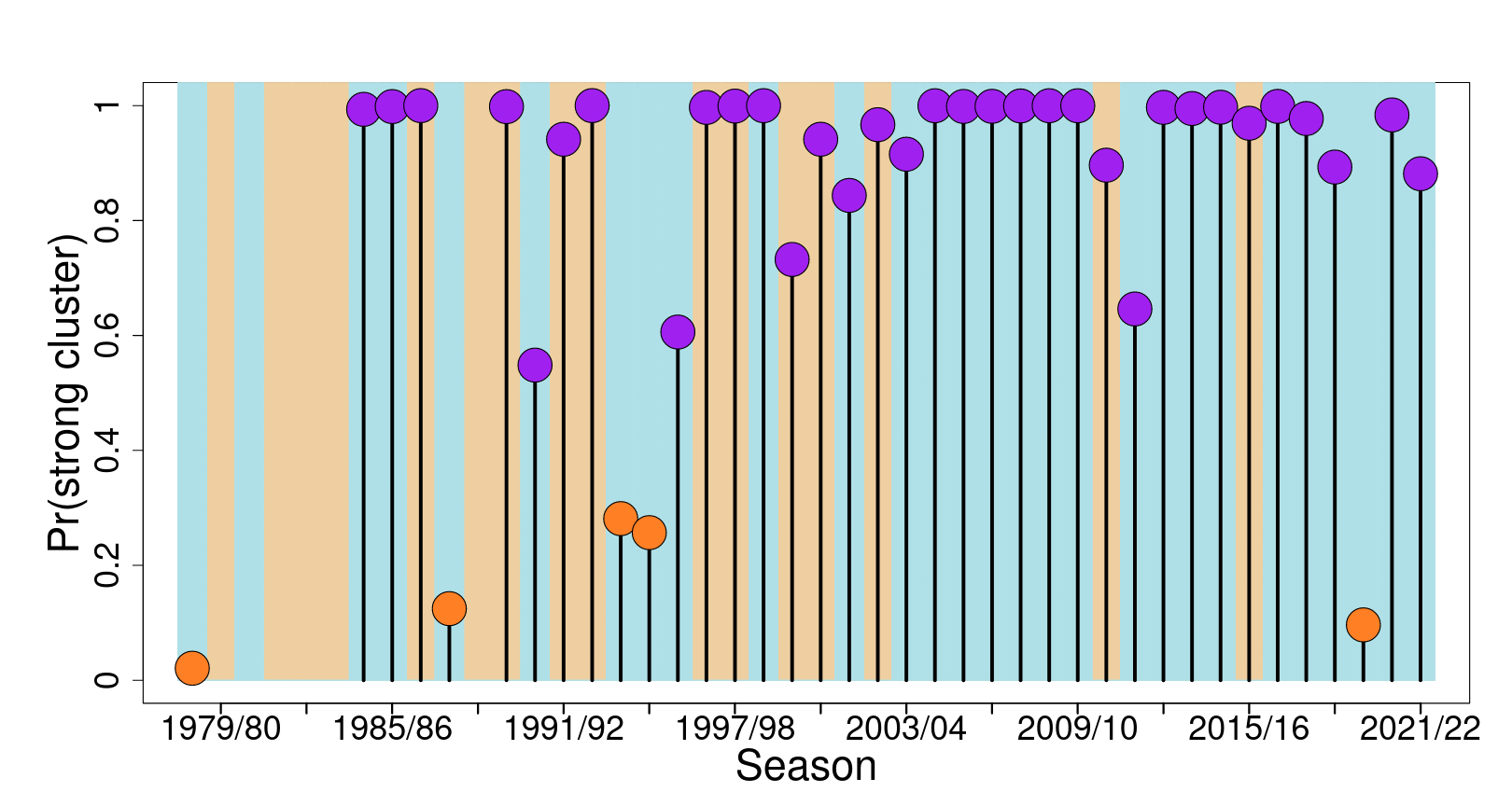}}{
					\caption{Chelsea.}}
			\end{subfloatrow}
						\begin{subfloatrow}[2]
				\ffigbox[\FBwidth]{
					\includegraphics[scale=0.15]{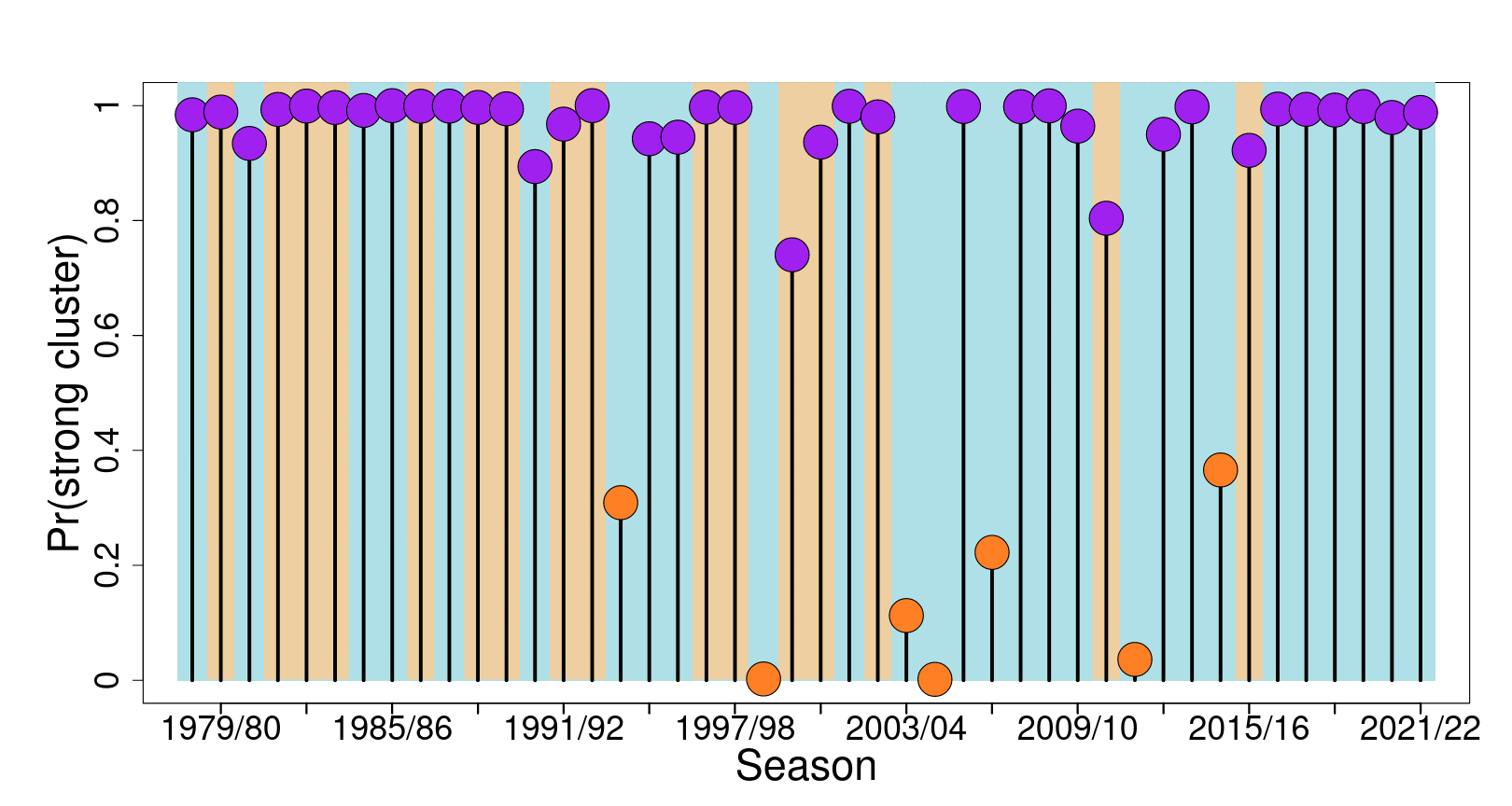}}{
					\caption{Liverpool.}}
				\ffigbox[\FBwidth]{
					\includegraphics[scale=0.15]{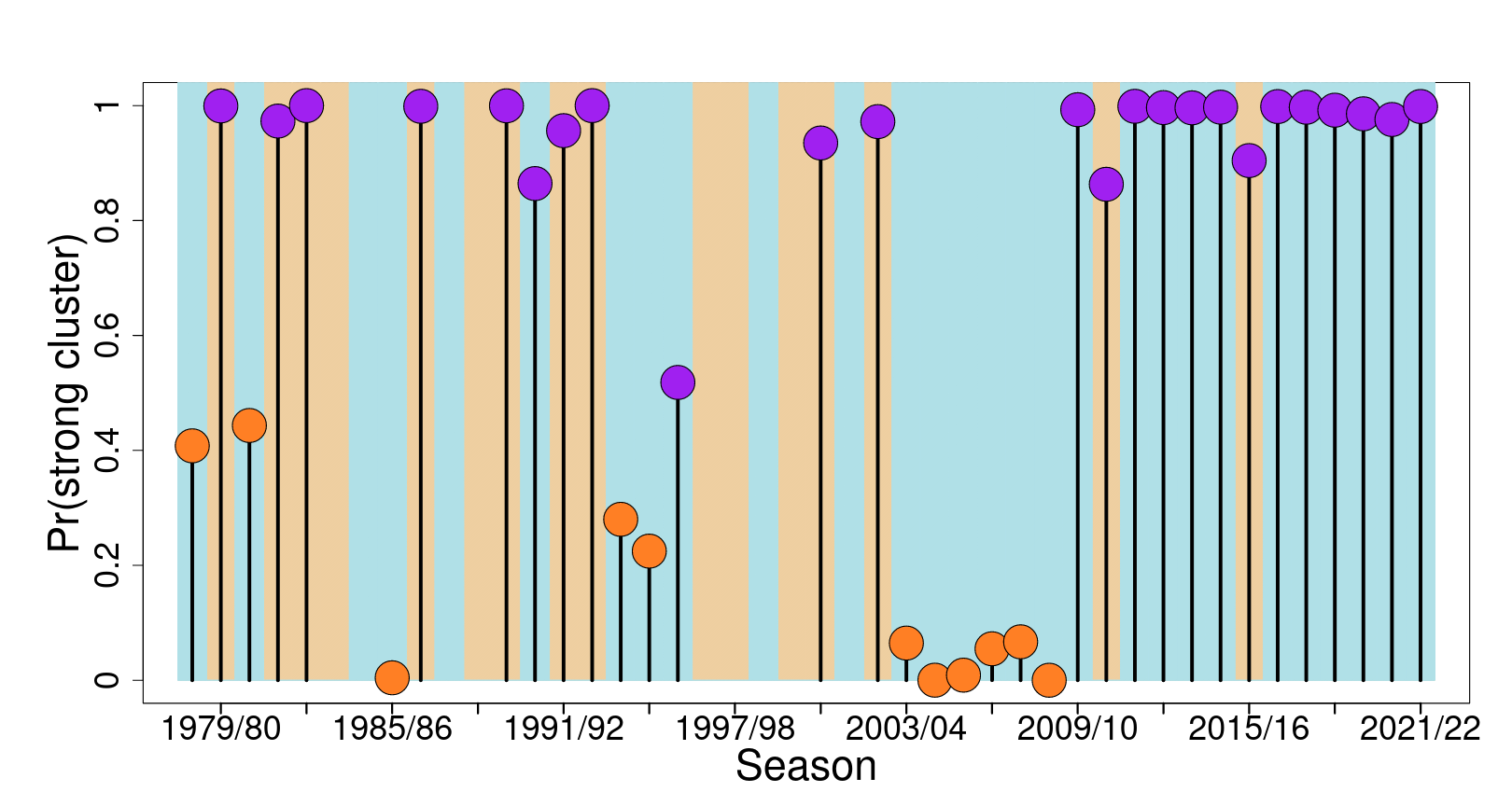}}{
					\caption{Manchester City.}}
			\end{subfloatrow}
			\begin{subfloatrow}[2]
				\ffigbox[\FBwidth]{%
					\includegraphics[scale=0.15]{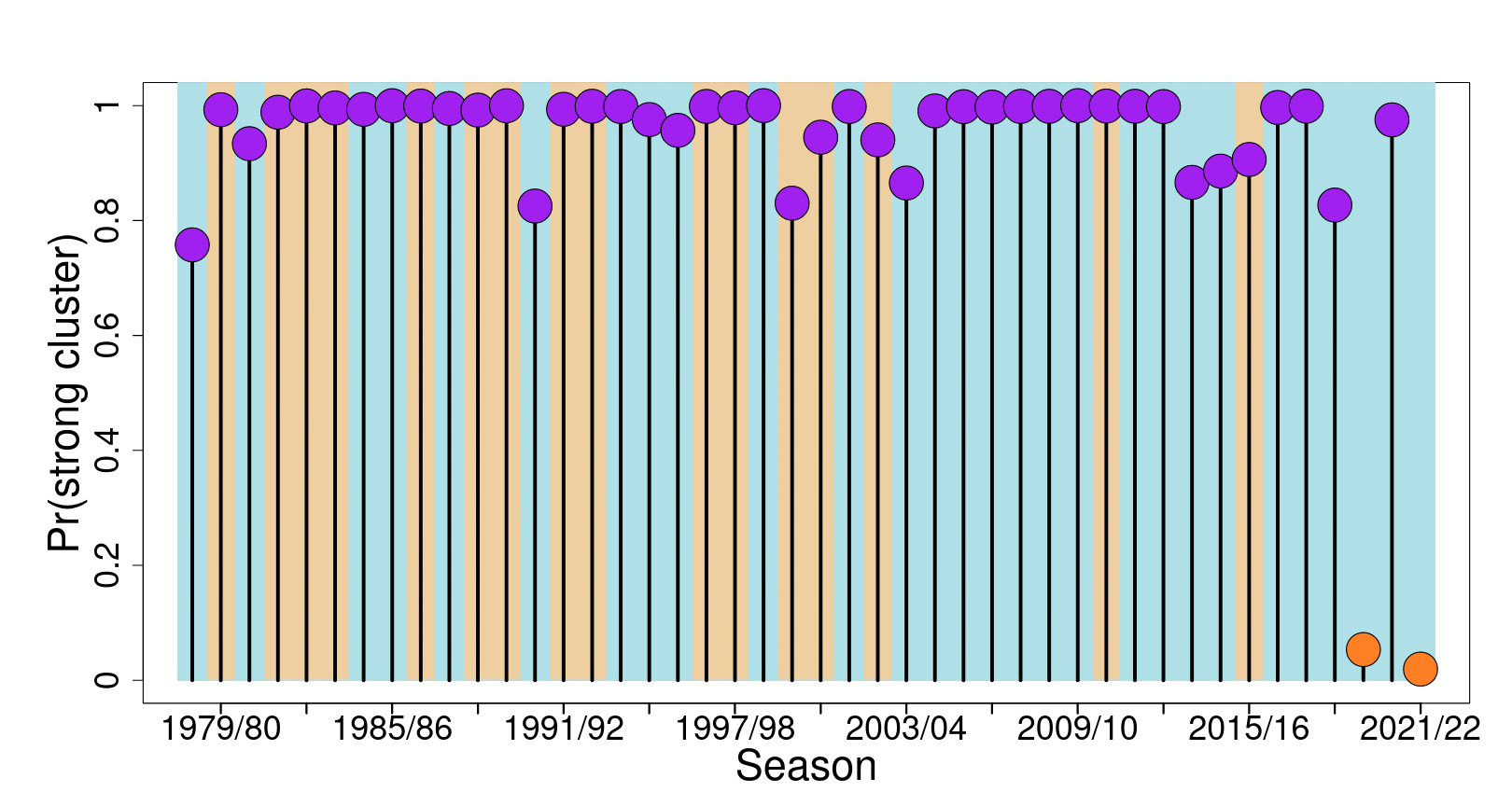}}{
					\caption{Manchester United.}}
				\ffigbox[\FBwidth]{%
					\includegraphics[scale=0.15]{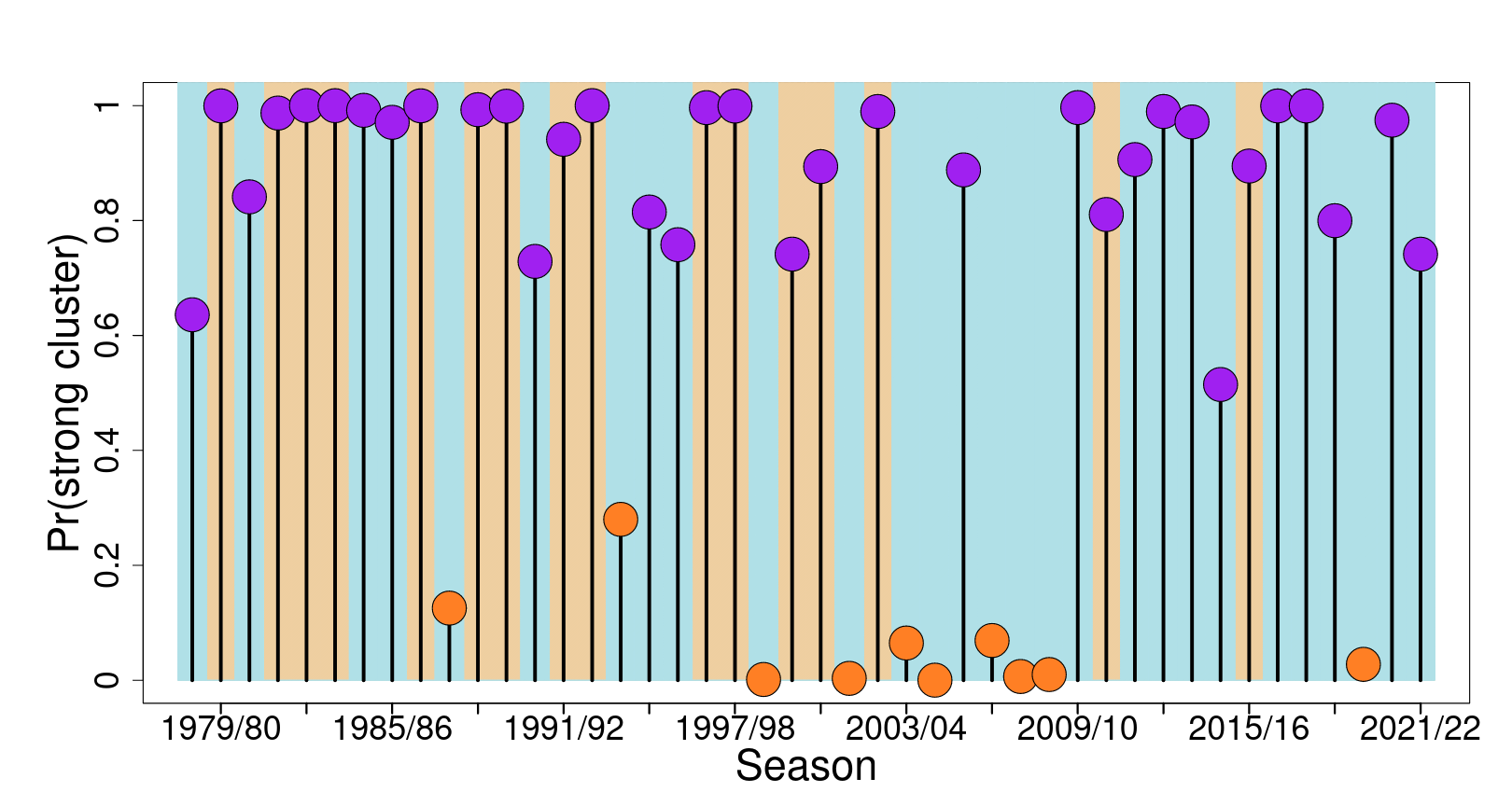}}{
					\caption{Tottenham Hotspur.}}
			\end{subfloatrow}
		}{
			\caption{Plot of posterior probability of allocation to the strongest block for some selected teams over the entire study period.
			The x-axis indicates each season in chronological order, while the y-axis displays the posterior probability of inclusion in the strongest block  where each team is indicated with a purple dot if its posterior allocation probability is greater than 0.5 or with an orange one otherwise. If no dot is present for any time period it means that team was not playing in the First division/Premier League for that season. The background indicates how many blocks have been chosen with highest posterior probability. In particular, a sand coloured background is used for single-block model and a light blue colour is for a two block model.\\
			}
			\label{fig:ProTEAMS}
		}
\end{figure}

We supplement more detail to the analysis above in Figure \ref{fig:ProTEAMS} by displaying the posterior probability of allocation to the strongest block for each season for Arsenal, Chelsea, Liverpool, Manchester city, Manchester United, Tottenham Hotspur over the entire course of the study. This provides additional information by illustrating that the posterior probability of allocation to the strongest team can vary considerably for each team and over each season. For example, in $2017/18$ there is much less support for Arsenal's inclusion in the strongest block (the estimated probability in this case turns out to be $0.63$). 

\subsubsection{Validation of the SBM approach using two standard indices}

\begin{figure}
 \centering
 
 \begin{tabular}{cc}
  \includegraphics[scale=0.32]{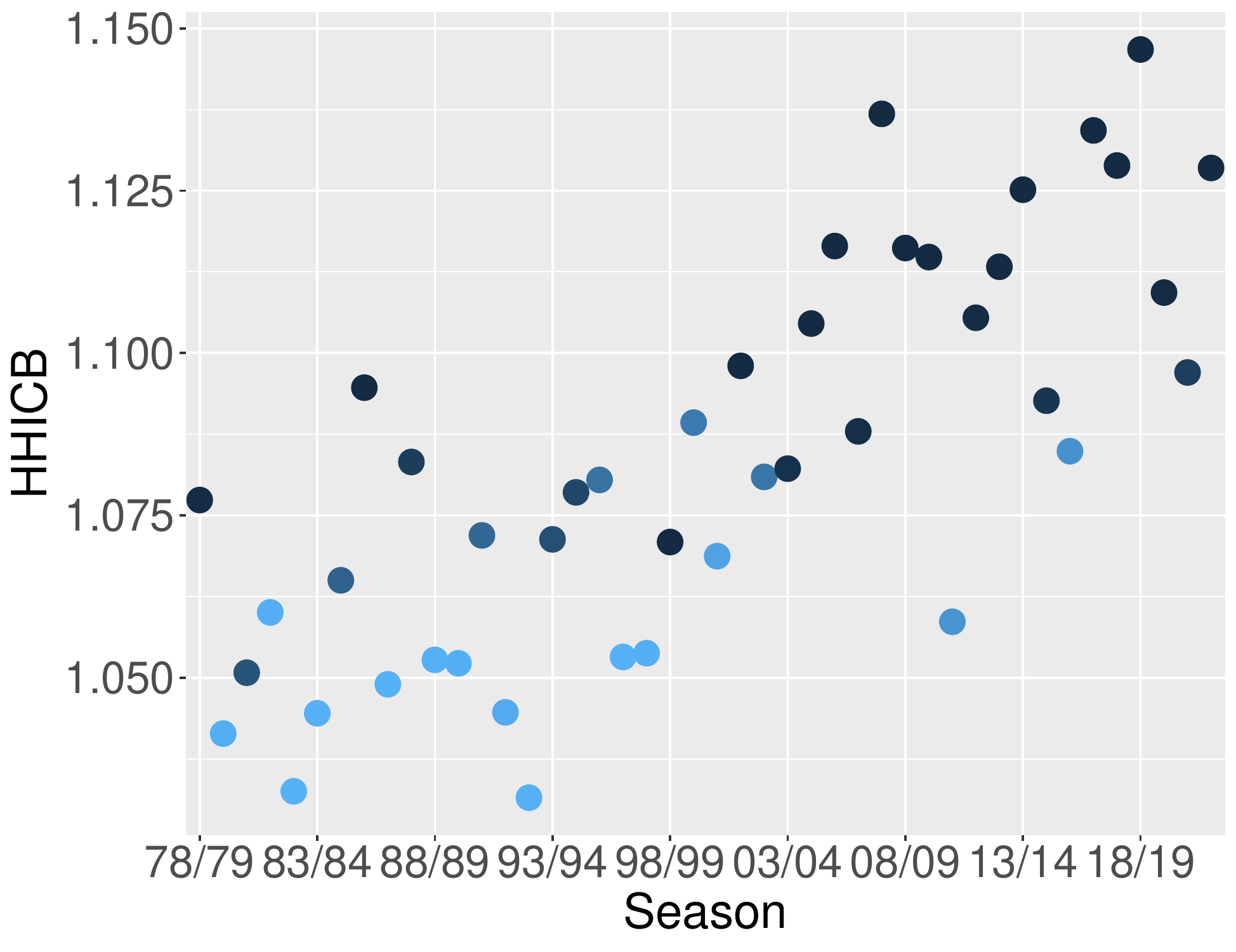} &
   \includegraphics[scale=0.32]{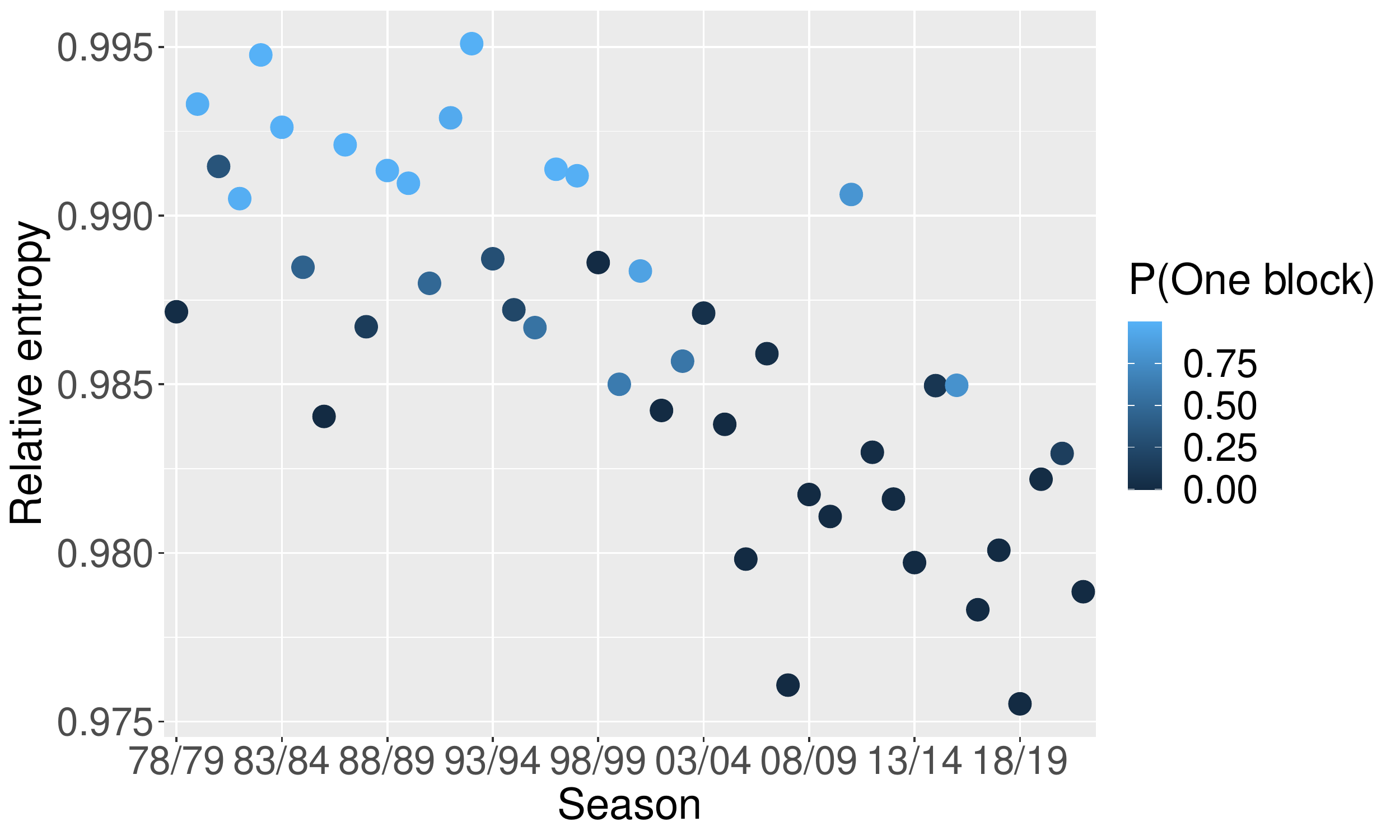} \\
   (a) & (b) \\
 \end{tabular}

 \caption{(a) The Herfindahl--Hirschman index of competitive balance (HHICB) and (b) the relative entropy are plotted for each season, while the posterior probability of a one block model, $\pi(K=1|\bm{y})$, is overlaid for each season. 
 Both plots illustrates that, in general, seasons with a lower HHICB score (higher relative entropy, respectively) tend to correspond to seasons where a single block had the most posterior support. 
}
 \label{fig:hicb_block}
\end{figure}

Here we revisit the statistics outlined in Section~\ref{sec:comp_balance}. We once again display both HHICB and relative entropy for each season. But now, in Figure~\ref{fig:hicb_block} we overlay the posterior probability of $K=1$ for each season. We see both plots illustrate that, in general, seasons with a lower HHICB score (higher relative entropy, respectively) tend to correspond to seasons where a single block had the most posterior support. Effectively, the HHICB and relative entropy scores are broadly in agreement with the results of our stochastic block model and provide a validation of our SBM approach. 

\subsection{Evidence for the emergence of a \textit{big-six} groups of teams}
\label{sec:topsix}

In this section we explore if there is some statistical evidence for two well known phenomena in popular discourse: the dominance of the \textit{big-four} (a group of four teams, namely, Arsenal, Chelsea, Liverpool and Manchester United) during the mid-2000's and the emergence of a \textit{big-six} from 2010 with the addition of Manchester City and Tottenham Hotspur. 
To investigate this, consider Table~\ref{tab:topblock} where we present the posterior allocation of teams to the strongest block of teams for the past two decades from $2003/04$ to $2021/22$ following Section~\ref{sec:top_block}.

\paragraph{From $2002/03$ to $2008/09$:}
Analysing Table~\ref{tab:topblock} from $2002/03$ to $2008/09$ indicates that Arsenal, Chelsea and Manchester United were ever present in the top block over this time period, while Liverpool were present in four of these seasons. This gives some credence to the notion of a \textit{big-four} block of teams during this period.

\paragraph{From $2009/10$ to $2021/22$:}
By contrast, season $2009/10$ onwards, signified the emergence of Manchester City and Tottenham Hotspur in the top block. In fact, $2009/10$ is when the former was purchased by the Abu Dhabi United Group. Since then, both teams have consistently been allocated to the top block of teams with the exception of Tottenham Hotspur in $2019/20$.
In terms of the period from $2009/10$ to $2021/22$, the collection of teams in the strongest block has been remarkable consistent until $2018/19$. However, over the past three seasons, Arsenal and Manchester United have been absent from the strongest block. While Chelsea have been absent in $2019/20$. This tentatively suggests that Arsenal and Manchester United may no longer be considered to among the strongest block of teams and that there is a re-emergence of a \textit{top-four}. Of course, more seasons are needed to validate this.  
Finally, it is worth noting that there are some seasons where some teams appear for a couple of consecutive seasons in the strongest block, for example, Everton from $2007/08$ to $2009/10$ and again from $2012/13$ to $2013/14$. We also remark that $2020/21$ was an exceptional season in the sense that the number of teams in the strongest block contained $15$ teams. This season was, of course, impacted by the COVID-19 pandemic when the majority of matches were played behind closed doors and the fact that the strongest block was so large, may suggest that the absence of crowds had the effect of increasing competitive balance.

\begin{table}
\begin{tabular}{c|cccccccc}
\toprule
& Arsenal & Chelsea & Liverpool & Man City & Man Utd & Tottenham & \multicolumn{2}{c}{Additional teams} \\
\rowcolor{navajowhite2}{\cellcolor{navajowhite2}02/03} &  \cmark & \cmark & \cmark &  & \cmark & \cmark & & \\
03/04 & \cmark & \cmark& \xmark& \xmark& \cmark& \xmark && \\ 
04/05 & \cmark & \cmark& \xmark& \xmark& \cmark& \xmark && \\
05/06 & \cmark & \cmark& \cmark& \xmark& \cmark& \cmark & \multicolumn{2}{c}{Blackburn, Newcastle} \\
06/07 & \cmark & \cmark& \xmark& \xmark& \cmark& \xmark && \\
07/08 & \cmark & \cmark& \cmark& \xmark& \cmark& \xmark &\multicolumn{2}{c}{Everton} \\
08/09 & \cmark & \cmark& \cmark& \xmark& \cmark& \xmark &\multicolumn{2}{c}{Everton, Aston Villa} \\
09/10 & \cmark & \cmark& \cmark& \cmark& \cmark& \cmark &\multicolumn{2}{c}{Everton, Aston Villa} \\
\rowcolor{navajowhite2}{\cellcolor{navajowhite2}10/11} & \cmark & \cmark & \cmark & \cmark & \cmark & \cmark & & \\
11/12 & \cmark & \cmark& \xmark& \cmark& \cmark& \cmark & \multicolumn{2}{c}{Newcastle} \\
12/13 & \cmark & \cmark& \cmark& \cmark& \cmark& \cmark &\multicolumn{2}{c}{Everton} \\
13/14 & \cmark & \cmark& \cmark& \cmark& \cmark& \cmark &\multicolumn{2}{c}{Everton}  \\
14/15 & \cmark & \cmark& \xmark& \cmark& \cmark& \cmark && \\
16/17 & \cmark & \cmark& \cmark& \cmark& \cmark& \cmark && \\
\rowcolor{navajowhite2}{\cellcolor{navajowhite2}15/16} & \cmark & \cmark & \cmark & \cmark & \cmark & \cmark & & \\
17/18 & \cmark & \cmark& \cmark& \cmark& \cmark& \cmark && \\
18/19 & \cmark & \cmark& \cmark& \cmark& \cmark& \cmark && \\
19/20 & \xmark & \xmark& \cmark& \cmark& \xmark& \xmark && \\
20/21 & \cmark & \cmark& \cmark& \cmark& \cmark& \cmark &\multicolumn{2}{c}{$9$ additional teams} \\
21/22 & \xmark & \cmark& \cmark& \cmark& \xmark& \cmark && \\
\bottomrule
Total no. of Seasons in & & & & & & & \\
the top block (out of 20) & 17/20 & 19/20 & 15/20 & 13/20 & 17/20 & 14/20 &  \\
\bottomrule
\end{tabular}
\caption{Posterior allocation of teams to the strongest block for each season since $2003/04$. Here the symbols \cmark and \xmark\; denote allocation or not, respectively, to the strongest block. Seasons with the highest probability of a single block model are highlighted in colour. Additional teams which were allocated to the strongest block are also listed per season. 
}
\label{tab:topblock}
\end{table}

\section{Conclusions}
\label{sec:conclusion}
In this paper we have developed a stochastic block model to allow a probabilistic assessment of competitive balance in the English football league. 
The important aspect of this model is that it allows one to assess the likely number of blocks of teams in a league, where a block consists of a number of teams such that the probability mass function of the outcome (a win, draw, or loss) is estimated to be the same for a match involving any two teams in that block. Similarly, a match involving a team from one block playing against any team from another block has its own between-block probability mass function for the outcome of that match. In contrast to previous approaches, our model approach yields a richer understanding of the nature of the competitiveness of a league through estimation of the posterior probability of the number of blocks, together with the most likely allocation of teams to each block. But also it allows one to estimate the number of teams allocated to the strongest block of teams, by integrating over the posterior uncertainty in the number of blocks.
    
In terms of our analysis of the English Premier League, we have uncovered evidence that the league was quite balanced from around $1980$ to $2003$. However, subsequent to that, there is strong evidence that the league has become more imbalanced since from $2003/04$ we see an emergence of seasons where two blocks are most probable, a posteriori. This is further supported by our analysis which shows that the estimated number of teams in the strongest block of teams is typically quite large from $1980$ to around $2003$, in contrast to the second half of the study period where the number of teams estimated to be in strongest block is, in general, much smaller. Our findings may serve to provide the Premier League with quantitative evidence of a lack of competitiveness over the past two decades which in turn may inform policy around potential reform of the league structure to address this issue.
For example, it would be of interest to understand the effect which a more equal redistribution of television revenue may have on competitive balance.

This paper could be extended in several directions. For example, the stochastic block model does not directly model for the number of goals scored by either team. In fact, there is a steady literature on statistical models for football match data beginning with \citep{dixoncoles97}, where a Poisson GLM framework is used to model the number of goals scored by either team. This has been extended by several authors, including \cite{karlis_ntzoufras03} to the bivariate Poisson setting. While \cite{rue2000prediction} extend the Dixon and Coles framework to a Bayesian dynamic GLM setting. There would therefore be interest in extending our modified stochastic block model to accommodate models for the number of goals scored by both teams using some of these frameworks. 
A further extension would be the development of a statistical changepoint model to assess the uncertainty around the season where a changepoint has occured. Finally we note that it would be natural to extend our SBM to incorporate temporal dependence across seasons.

\section*{Acknowledgements} 
The Insight Centre for Data Analytics is supported by Science Foundation Ireland under Grant Number 12/RC/2289$\_$P2. 

\section*{Datasets and code}
All of the datasets and \texttt{R} code used to reproduce the results and figures in this paper can be found at \url{https://github.com/francescabasini/Football_SBM} \\

\bibliographystyle{chicago}
\bibliography{references}

\newpage

\newpage
\begin{appendices}
\section{Calculation of the collapsed SBM}

\subsection{Collapsing p}
\begin{center}
	\begin{adjustbox}{max width=\textwidth}
		\parbox{\linewidth}{
		\begin{align}  
		f(\bm{y}|\bm{z},K)&=\int_{\bm{p}}f(\bm{y}|\bm{z},\bm{p},K)\pi(\bm{p}|K)d\bm{p} \nonumber \\
		&=\int_{p_\omega^{kl}\in \underline{p}^{kl} \in \bm{p}} \prod_{i=1}^{N-1}\prod_{\underset{j\neq i}{j=1}}^N \prod_{k=1}^K\prod_{l=1}^K \left(\prod_{\omega=1}^3 \left(p_\omega^{kl}\right)^{I(y_{ij}=\omega)} \right)^{I\left(z_i=k\right)I\left(z_j=l\right)}\cdot \prod_{k=1}^K\prod_{l=1}^K \frac{1}{B(\bm{\beta})} \prod_{\omega=1}^3 \left( p_\omega^{kl} \right) ^{\beta_\omega-1}d p_\omega^{kl} \nonumber \\
		&\begin{aligned}= \int_{p_\omega^{kl}\in \underline{p}^{kl} \in \bm{p}} \prod_{k=1}^K\prod_{l=1}^K \prod_{\omega=1}^3 \left(p_\omega^{kl}\right)^{\sum_{i=1}^{N-1}\sum_{\underset{j\neq i}{j=1}}^N I(y_{ij}=\omega){I\left(z_i=k\right)I\left(z_j=l\right)}} 
                 \prod_{k=1}^K\prod_{l=1}^K \frac{1}{B(\bm{\beta})} \prod_{\omega=1}^3 \left( p_\omega^{kl} \right) ^{\beta_\omega-1}d p_\omega^{kl} \end{aligned} \label{eqn:appendix_collapse}
		\end{align}
	}
	\end{adjustbox}
\end{center}
Here we define 
\begin{equation}\label{eq:N_kl}
N_{kl}^{\omega} = \sum_{i=1}^{N-1}\sum_{\underset{j\neq i}{j=1}}^N {I(y_{ij}=\omega) I\left(z_i=k\right)I\left(z_j=l\right)}, 
\end{equation}
for $k=1,\dots,K$ and $l=1,\dots,K$, so that $N_{kl}^{\omega}$ counts the number of times that the outcome $\omega$ was observed for all games involving a team allocated to block $k$ playing at home against a team allocated to block $l$. 
Thus, we can rewrite (\ref{eqn:appendix_collapse}) as:
\begin{align}
	f(\bm{y}|\bm{z},K) &= \int_{p_\omega^{kl}\in \underline{p}^{kl} \in \bm{p}} \prod_{k=1}^K\prod_{l=1}^K \prod_{\omega=1}^3  \left(p_\omega^{kl}\right)^{N_{kl}^{\omega} } \cdot \prod_{k=1}^K\prod_{l=1}^K \frac{1}{B(\bm{\beta})} \prod_{\omega=1}^3 \left( p_\omega^{kl} \right) ^{\beta_\omega-1}d p_\omega^{kl}. \nonumber \\
	=& \prod_{k=1}^K\prod_{l=1}^K \frac{1}{B(\bm{\beta})} \int_{p_\omega^{kl}\in \underline{p}^{kl} \in \bm{p}} \prod_{\omega=1}^3\left(p_\omega^{kl}\right)^{N_{kl}^{\omega} +\beta_\omega-1}  d p_\omega^{kl} \label{eqn:collapseP},
\end{align}
Inside (\ref{eqn:collapseP}), we recognise the kernel of a 3-dimensional Dirichlet distribution of the form:
$$(p_1^{kl},p_2^{kl},p_3^{kl}) \sim Dir( N_{kl}^{\omega=1}+ \beta_1, N_{kl}^{\omega=2}+ \beta_2, N_{kl}^{\omega=3}+ \beta_3).$$
Therefore, we can re-write the right hand side of equation (\ref{eqn:collapseP}) as: 
$$ \prod_{k=1}^K\prod_{l=1}^K \frac{1}{B(\bm{\beta})}
	\frac{\prod_{\omega=1}^3\Gamma( N_{kl}^{\omega}+\beta_\omega)}{\Gamma(\sum_{\omega=1}^3 \left( N_{kl}^{\omega}+\beta_\omega)\right)},$$
where
	$$\frac{1}{B(\bm{\beta})} =\frac{\Gamma(\sum_{\omega=1}^3 \beta_{\omega})}{\prod_{\omega=1}^3 \Gamma(\beta_{\omega})}.$$
This yields the expression
\begin{equation*}
	f(\bm{y}|\bm{z},K)= \prod_{k=1}^K\prod_{l=1}^K \frac{\Gamma(\sum_{\omega=1}^3 \beta_{\omega})}{\prod_{\omega=1}^3 \Gamma(\beta_{\omega})} \frac{\prod_{\omega=1}^3\Gamma( N_{kl}^{\omega}+\beta_{\omega})}{\Gamma(\sum_{\omega=1}^3 \left( N_{kl}^{\omega}+ \beta_{\omega})\right) }.
\end{equation*}
Recall that in our framework the concentration parameters of the prior Dirichlet distribution of $\underline{p}^{kl} \text{ for } k,l=1,\ldots, K$ are all set to be equal to have a uniform prior,
$$\beta_\omega =1, \;\;\mbox{for}\;\; \ \omega=1,2,3 \ ,$$ 
resulting in,
\begin{equation}
	f(\bm{y}|\bm{z},K)= \prod_{k=1}^K\prod_{l=1}^K \Gamma(3) \frac{\prod_{\omega=1}^3\Gamma( N_{kl}^{\omega}+1)}
	{\Gamma(\sum_{\omega=1}^3 \left( N_{kl}^{\omega} + 1)\right)}. \label{eqn:collapseP_final}
\end{equation}

\subsection{Collapsing theta }
\begin{align}
	\pi(\bm{z}|K)&=\int_{\bm{\Theta}}\pi(\bm{z}|\bm{\theta},K)\pi(\bm{\theta}|K)d\bm{\Theta} \nonumber\\
	&=\int_{\bm{\Theta}}\prod_{i=1}^N Multi(z_i;1,\bm{\theta}) \cdot Dir(\bm{\theta}; \bm{\gamma})d\bm{\Theta}\nonumber\\
	&=\int_{\bm{\Theta}}\prod_{i=1}^N \prod_{k=1}^K\theta_k^{I\left(z_i=k\right)} \cdot \frac{\Gamma(\sum_{k=1}^K\gamma_k)}{\prod_{k=1}^K \Gamma\left(\gamma_k\right)} \prod_{k=1}^K\theta_k^{\gamma_k-1}d\bm{\Theta}\nonumber\\
	&=\Gamma(\sum_{k=1}^K\gamma_k)\int_{\bm{\Theta}} \prod_{k=1}^K\theta_k^{\sum_{i=1}^NI\left(z_i=k\right)}\frac{1}{\prod_{k=1}^K\Gamma(\gamma_k)}\prod_{k=1}^K\theta_k^{\gamma_k-1}d\bm{\Theta}\nonumber\\
	&=\Gamma(\sum_{k=1}^K\gamma_k)\int_{\bm{\Theta}} \prod_{k=1}^K\theta_k^{\sum_{i=1}^NI\left(z_i=k\right)+\gamma_k-1}\frac{1}{\Gamma(\gamma_k)}d\bm{\Theta} \label{eq:CollapsedTheta}
\end{align}
We set 
\begin{equation}\label{eq:nk}
	n_k=\sum_{i=1}^N{I\left(z_i=k\right)}, \ k=1,\ldots, K,
\end{equation}
where each $n_k$ accounts for the number of nodes/teams allocated to block $k$.
Therefore we can rewrite the right hand side of (\ref{eq:CollapsedTheta}) as:
\begin{align*}
	&\Gamma(\sum_{k=1}^K\gamma_k)\int_{\bm{\Theta}} \prod_{k=1}^K\theta_k^{n_k+\gamma_k-1}\frac{1}{\Gamma(\gamma_k)}d\bm{\Theta}\nonumber\\
	&=\frac{\Gamma(K\cdot \gamma_0)}{\prod_{k=1}\Gamma(\gamma_0)}\int_{\bm{\Theta}} \prod_{k=1}^K\theta_k^{n_k+\gamma_0-1}d\bm{\Theta} \ .
\end{align*}
We recognise the integral above as the density of a $Dir(n_1+\gamma_0, \ldots, n_K+\gamma_0)$ distribution, thus we can write:
\begin{align*}
	\pi(\bm{z}|K)&=\frac{\Gamma(K\cdot \gamma_0)}{\prod_{k=1}^K\Gamma(\gamma_0)} \frac{\prod_{k=1}^K\Gamma(n_k+\gamma_0)}{\Gamma(\sum_{k=1}^K (n_k+\gamma_0))}\\
	&=\frac{\prod_{k=1}^K \Gamma(n_k+\gamma_0)}{\Gamma^K(\gamma_0)} \frac{\Gamma(K\cdot \gamma_0)}{\Gamma(N+K\gamma_0)}\ . \nonumber
\end{align*}
Setting the hyperparameter $\gamma_0=1$, leads to
\begin{equation}
	\pi(\bm{z}|K)=\prod_{k=1}^K \Gamma(n_k+1) \frac{\Gamma(K)}{\Gamma(N+K)}. \label{eqn:collapse_theta}
\end{equation}
	
\subsubsection*{Collapsed posterior}
Using (\ref{eqn:collapseP_final}) and (\ref{eqn:collapse_theta}) results in the expression for the collapsed posterior,
\begin{align}
	\pi(\bm{z}|\bm{y},K) &\propto f(\bm{y}|\bm{z},K)\pi(\bm{z}|K) \pi(k) \nonumber \\
	&=\prod_{k=1}^K\prod_{l=1}^K \Gamma(3) \frac{\prod_{\omega=1}^3\Gamma( N_{kl}^{\omega}+1)}{\Gamma(\sum_{\omega=1}^3 \left( N_{kl}^{\omega} +1)\right)} \cdot \prod_{k=1}^K\Gamma(n_k+1)  \frac{\Gamma(K)}{\Gamma(N+K)} \times \frac{1}{K!} \, .
\end{align}.



\section{Details of the MCMC move types} 

\subsection{Acceptance probabilities of the MK move}
\subsubsection*{Insert attempt}
The proposal probabilities formulated in terms of an \textit{insert} attempt are:
\begin{align*}
P_{prop}((\bm{z},K)\rightarrow (\bm{z}',K')) &= P_{prop}((\bm{z},K)\rightarrow (\bm{z},K+1)) \\
& = Pr(\text{Insert an empty cluster})\\
&= \begin{cases}
0.5, \ \text{if }K < K_{max}\\
0, \text{ if } K=K_{max}.
\end{cases}\\
P_{prop}((\bm{z}',K')\rightarrow (\bm{z},K)) &= P_{prop}((\bm{z},K+1)\rightarrow (\bm{z},K)) \\
&=Pr(\text{Delete an empty cluster})\\
&= 0.5
\end{align*}
Recall that the algorithm will reject the move to increase the number of clusters when $K=K_{max}$ and also that this move does not propose to change $\bm{z}$. 
Here, when $K<K_{max}$ the ratio of the posterior density at the proposed and current states can be written as:
\begin{align}
\frac{\pi(\bm{z},K'|\bm{y})}{\pi(\bm{z},K|\bm{y})}&= \frac{f(\bm{y}|\bm{z})\pi(\bm{z}|K')\pi(K')}{f(\bm{y}|\bm{z})\pi(\bm{z}|K)\pi(K)}\nonumber\\
&= \frac{\pi(\bm{z}|K+1)}{\pi(\bm{z}|K)}\times \frac{\pi(K+1)}{\pi(K)} \nonumber\\
&= \frac{\left( \prod_{k=1}^{K+1}\Gamma(n_{k}+1)  \frac{\Gamma((K+1)) }{\Gamma(N+(K+1))}\right)} 
        {\left(\prod_{k=1}^K\Gamma(n_k+1) \frac{\Gamma(K)}{\Gamma(N+K)}\right)}\times \frac{K!}{(K+1)!} \ . \label{eq:InsMassRatio}
\end{align}
Note that since the newly inserted cluster is not allocated any nodes,
\[
 \prod_{k=1}^{K+1}\Gamma(n_{k}+1) = \prod_{k=1}^K\Gamma(n_k+1)
\]
and so (\ref{eq:InsMassRatio}) reduces to
\begin{align} \frac{\pi(\bm{z},K'|\bm{y})}{\pi(\bm{z},K|\bm{y})} &= \frac{\left( \frac{\Gamma((K+1)}{\Gamma(N+(K+1)}\right)}{\left( \frac{\Gamma(K)}{\Gamma(N+K)}\right)}\times \frac{1}{K+1} \nonumber \\
&= \frac{K}{(N+K)(K+1)}.
\label{eq:InsMassRatio_simple}
\end{align}
Since we have symmetric proposal probabilities when $K<K_{max}$, the acceptance probability of the \textit{insert attempt} is:
	\begin{equation} 
	\alpha = \frac{K}{(N+K)(K+1)} \label{eq:alphaInsert}
	\end{equation}

\subsubsection*{Delete attempt}
For the \textit{delete attempt}, provided that $K>1$, the proposal probabilities are:

\begin{align*}
P_{prop}((\bm{z},K)\rightarrow (\bm{z}',K')) &= P_{prop}((\bm{z},K)\rightarrow (\bm{z},K-1)) \\
& = Pr(\text{delete an empty cluster})\\
&= \begin{cases}
0.5, \ \text{if }K >1 \\
0, \text{ if } K=1.
\end{cases}\\
P_{prop}((\bm{z}',K')\rightarrow (\bm{z},K)) &= P_{prop}((\bm{z},K-1)\rightarrow (\bm{z},K)) \\
&=Pr(\text{Insert an empty cluster})\\
&= 0.5
\end{align*}
Notice again that, due to the way the algorithm is built, we will prevent a \textit{delete} move from taking place if $K=1$. 
Here, when $K>1$ the ratio of the posterior density at the proposed and current states can be written as:
\begin{align}
\frac{\pi(\bm{z},K'|\bm{y})}{\pi(\bm{z},K|\bm{y})}&= \frac{f(\bm{y}|\bm{z})\pi(\bm{z}|K')\pi(K')}{f(\bm{y}|\bm{z})\pi(\bm{z}|K)\pi(K)}\nonumber\\
&= \frac{\pi(\bm{z}|K-1)}{\pi(\bm{z}|K)}\times \frac{\pi(K-1)}{\pi(K)} \nonumber\\
&= \frac{\left( \prod_{k=1}^{K-1}\Gamma(n_{k}+1)  \frac{\Gamma((K-1)) }{\Gamma(N+(K-1))}\right)} 
        {\left(\prod_{k=1}^{K}\Gamma(n_k+1) \frac{\Gamma(K)}{\Gamma(N+K}\right)}\times \frac{K!}{(K-1)!} \ .  \label{eq:AddMassRatio}
\end{align}
Since a \textit{delete} move involves removing an empty cluster,
\[
 \prod_{k=1}^{K-1}\Gamma(n_{k}+1) = \prod_{k=1}^K\Gamma(n_k+1)
\]
and so (\ref{eq:AddMassRatio}) reduces to 
\[\frac{\pi(\bm{z},K'|\bm{y})}{\pi(\bm{z},K|\bm{y})} = \frac{K(N+K-1)}{K-1} \, .\]
The acceptance probability of \textit{delete} attempt when $K>1$, since this results in symmetric proposal probabilities, is:
\begin{equation}
\alpha = \frac{K(N+K-1)}{K-1} .  \label{eq:alphaDelete}
\end{equation}

\subsection{Acceptance probability for Metropolis-within-Gibbs move}
Here we provide some details on the acceptance probability outlined in equation (11) in the manuscript. 
We can express this as 
\begin{equation}
 \frac{\pi(\bm{z}',K|\bm{y})}{\pi(\bm{z}^{(s+1)},K|\bm{y})} =  \frac{ \prod_{k=1}^K\prod_{l=1}^K  \frac{\prod_{\omega=1}^3\Gamma( N_{kl}^{\omega'}+1)}{\Gamma(\sum_{\omega=1}^3 \left( N_{kl}^{\omega'} +1)\right)}}{ \prod_{k=1}^K\prod_{l=1}^K  \frac{\prod_{\omega=1}^3\Gamma( N_{kl}^{\omega}+1)}{\Gamma(\sum_{\omega=1}^3 \left( N_{kl}^{\omega} +1)\right)} } \times 
 \frac{\prod_{k=1}^K\Gamma(n_k^{'}+1)}{\prod_{k=1}^K\Gamma(n_k^{'}+1)},
 \label{eqn:m-within-g-ratio}
\end{equation}
where $N_{kl}^{\omega'}$ is the equivalent quantity calculated in~(\ref{eq:N_kl}), but applied to the proposed allocation vector $\bm{z}'$,
\begin{equation*}
N_{kl}^{\omega'} = \sum_{i=1}^{N-1}\sum_{\underset{j\neq i}{j=1}}^N {I(y_{ij}=\omega) I\left(z_i^{'}=k\right)I\left(z_j^{'}=l\right)}.
\end{equation*}
However, recall that proposed allocation vector $\bf{z}'$ is identical to the current state of the allocation vector, $\bm{z}^{(s+1)}$, except for its $i$th element, $z_i'$, which we suppose is allocated the label $k_1$. Further, suppose that the current state $z_i^{(s+1)}=k_0$. It turns out that this leads to some simplification in the first term in the right hand side of~(\ref{eqn:m-within-g-ratio}), since in this case, 
\begin{equation}
N_{kl}^{\omega'}= \begin{cases}
N_{kl}^{\omega'}, \text{ if } k=k_0 \ or \ k=k_1 \ or \ l=k_1 \ or \ l =k_0,\\
N_{kl}^\omega \ , \ \text{otherwise.}
\end{cases} 
\end{equation}
This leads to a significant saving in the calculation of this term in~(\ref{eqn:m-within-g-ratio}). The second term on the right hand side of~(\ref{eqn:m-within-g-ratio}) can also be simplified. Since in this case, we note that $n_k^{'}$ is defined similar to~(\ref{eq:nk}) as 
\begin{equation}
	n_k^{'}=\sum_{i=1}^N{I\left(z_i^{'}=k\right)}, \ k=1,\ldots, K,
\end{equation}
and again we remark that $\bm{z}'$ is identical to the current state of the allocation vector, $\bm{z}^{(s+1)}$, except for its $i$th element, $z_i'=k_1$. This implies that 
\begin{equation}
n_k^{'}= \begin{cases}
n_{k_1}+1, \text{ if } k=k_1, \\
n_{k_0}-1, \text{ if } k=k_0, \\
n_k \ , \ \text{otherwise.}
\end{cases} 
\end{equation}
consequently, the second term on the right hand side of~(\ref{eqn:m-within-g-ratio}) can be written as
\begin{align}
\frac{\prod_{k=1}^K\Gamma(n'_k+1)}{\prod_{k=1}^K\Gamma(n_k+1)}&= \frac{\Gamma(n_{k_0})\Gamma(n_{k_1}+2)}{\Gamma(n_{k_1}+1)\Gamma(n_{k_0}+1)}\nonumber\\
&=\frac{n_{k_1}+1}{n_{k_0}} \ . \label{eq:SecondTerm}
\end{align}

\subsection{Acceptance probabilities of the AE move}

\subsubsection{Ejection attempt}
The proposal probabilities of an \textit{ejection attempt} can be written as:
\begin{align}
P_{prop}((\bm{z}',K')\rightarrow (\bm{z},K)) &= P_{prop}((\bm{z}',K+1)\rightarrow (\bm{z},K))\nonumber\\
&=Pr(\text{\footnotesize absorb a cluster})\times \frac{1}{\substack{\text{ \# labels for} \\ \text{absorbing cluster}}}\times \frac{1}{\substack{\text{ \# labels for} \\ \text{absorbed cluster}}}\nonumber \\
&=(1-p_K^e)\frac{1}{K+1}\frac{1}{K}\nonumber . 
\end{align}
While the reverse proposal probability is detailed as:
\begin{align}
P_{prop}((\bm{z},K)\rightarrow (\bm{z}',K')) &= P_{prop}((\bm{z},K)\rightarrow (\bm{z}',K+1))\nonumber\\
&=Pr(\text{\footnotesize eject a cluster})\times \frac{1}{\substack{\text{ \# labels for}\\\text{ ejecting cluster} }} \times \frac{1}{\substack{\text{\# labels for} \\ \text{ejected cluster}}} \times Pr(\text{\footnotesize new allocation for }\bm{z})\nonumber \\
&=p_K^e\times \frac{1}{K+1}\times \frac{1}{K}\times \binom{n_{j_1}+n_{j_2}}{n_{j_1}} p_E^{n_{j_2}}(1-p_E)^{n_{j_1}} \cdot \frac{p_E^{a-1}(1-p_E)^{a-1}}{B(a,a)} \label{eq:eject2}
\end{align}
for $p_E \sim Beta(a,a)$ and 
where, 
\begin{align*}
n_{j_1}&=\sum_{i=1}^N I(z'_i=j_1) \text{ is the number of nodes in the ejecting cluster after reallocation,} \\
n_{j_2}&=\sum_{i=1}^N I(z'_i=j_2) \text{ is the number of nodes in the ejected cluster.}
\end{align*}

After integrating out with respect to the distribution of $p_E$, the resulting proposal probability in (\ref{eq:eject2}) reduces to:
\begin{align*}
&p_K^e\times \frac{1}{K+1}\times \frac{1}{K} \times \binom{n_{j_1}+n_{j_2}}{n_{j_1}}\frac{1}{B(a,a)}\int_0^1 p_E^{n_{j_2}+a-1}(1-p_E)^{n_{j_1}+a-1} d p_E \\
&=p_K^e\times \frac{1}{K+1}\times \frac{1}{K} \times \frac{(n_{j_1}+n_{j_2})!}{n_{j_1}!n_{j_2}!}\frac{\Gamma(2a)}{\Gamma(a)\Gamma(a)}\frac{\Gamma(n_{j_1}+a)\Gamma(n_{j_2}+a)}{\Gamma(n_{j_1}+n_{j_2}+2a)}\\
&=p_K^e\times \frac{1}{K+1}\times \frac{1}{K} \times \frac{1}{n_{j_1}+n_{j_2}+1}.
\end{align*}
The final equation above follows from setting $a=1$, so that $p_E\sim U(0,1)$, as in McDaid et al. (2013).

The posterior probability ratio for the \textit{ejection attempt} takes the following form:
\begin{align*} 
\frac{\pi(\bm{z}',K'|Y)}{\pi(\bm{z},K|Y)} &= \frac{f(Y|\bm{z}') \pi(\bm{z}'|K')\pi(K')}{f(Y|\bm{z}) \pi(\bm{z}|K)\pi(K)}\\
&=\frac{f(Y|\bm{z}') \pi(\bm{z}'|K+1)\pi(K+1)}{f(Y|\bm{z}) \pi(\bm{z}|K)\pi(K)}\\
&= \frac{ \prod_{k=1}^{K+1}\prod_{l=1}^{K+1}  \frac{\prod_{\omega=1}^3\Gamma( N_{kl}^{\omega'}+1)}{\Gamma(\sum_{\omega=1}^3 \left( N_{kl}^{\omega'} +1)\right)}}{ \prod_{k=1}^K\prod_{l=1}^K  \frac{\prod_{\omega=1}^3\Gamma( N_{kl}^{\omega}+1)}{\Gamma(\sum_{\omega=1}^3 \left( N_{kl}^{\omega} +1)\right)} } \times 
 \frac{\prod_{k=1}^{K+1}\Gamma(n_k^{'}+1) \frac{\Gamma(K+1)}{\Gamma(N+K+1)} }{\prod_{k=1}^K\Gamma(n_k+1) \frac{\Gamma(K)}{\Gamma(N+K)} } \times \frac{K!}{(K+1)!}\\
&=\frac{ \prod_{k=1}^{K+1}\prod_{l=1}^{K+1}  \frac{\prod_{\omega=1}^3\Gamma( N_{kl}^{\omega'}+1)}{\Gamma(\sum_{\omega=1}^3 \left( N_{kl}^{\omega'} +1)\right)}}{ \prod_{k=1}^K\prod_{l=1}^K  \frac{\prod_{\omega=1}^3\Gamma( N_{kl}^{\omega}+1)}{\Gamma(\sum_{\omega=1}^3 \left( N_{kl}^{\omega} +1)\right)} } \times 
 \frac{\prod_{k=1}^{K+1}\Gamma(n_k^{'}+1) }{\prod_{k=1}^K\Gamma(n_k+1)} \times \frac{K}{(N+K)(K+1)}
\end{align*}

\subsubsection{Absorption attempt}
The proposal probabilities of the \textit{absorption move} are:
\begin{align}
P_{prop}((\bm{z}',K')\rightarrow (\bm{z},K)) &= P_{prop}((\bm{z}',K-1)\rightarrow (\bm{z},K))\nonumber\\
&=Pr(\text{\footnotesize absorb a cluster})\times \frac{1}{\substack{\text{ \# labels for} \\ \text{absorbing cluster}}}\times \frac{1}{\substack{\text{ \# labels for} \\ \text{absorbed cluster}}}\nonumber \\
&=(1-p_K^e)\frac{1}{K-1}\frac{1}{K}\nonumber . 
\end{align}
\begin{align}
P_{prop}((\bm{z},K)\rightarrow (\bm{z}',K')) &= P_{prop}((\bm{z},K)\rightarrow (\bm{z}',K+1))\nonumber\\
&=Pr(\text{\footnotesize eject a cluster})\times \frac{1}{\substack{\text{ \# labels for}\\\text{ ejecting cluster} }} \times \frac{1}{\substack{\text{\# labels for} \\ \text{ejected cluster}}} \times Pr(\text{\footnotesize new allocation for }\bm{z})\nonumber \\
&=p_K^e\times \frac{1}{K-1}\times \frac{1}{K}\times \binom{n_{j_1}+n_{j_2}}{n_{j_1}} p_E^{n_{j_2}}(1-p_E)^{n_{j_1}} \cdot \frac{p_E^{a-1}(a-p_E)^{a-1}}{B(a,a)} \nonumber\\
&=p_K^e\times \frac{1}{K-1}\times \frac{1}{K}\times \frac{1}{n_{j_1}+n_{j_2}+1}\, . \nonumber 
\end{align}
Where, as before, the final equation above results from setting $a=1$.
The posterior probability ratio is given by:
\begin{align} \frac{\pi(\bm{z}',K'|Y)}{\pi(\bm{z},K|Y)} &= \frac{f(Y|\bm{z}') \pi(\bm{z}'|K')\pi(K')}{f(Y|\bm{z}) \pi(\bm{z}|K)\pi(K)} \nonumber\\
&=\frac{f(Y|\bm{z}') \pi(\bm{z}'|K-1)\pi(K-1)}{f(Y|\bm{z}) \pi(\bm{z}|K)\pi(K)}\nonumber \\
&=\frac{ \prod_{k=1}^{K-1}\prod_{l=1}^{K-1}  \frac{\prod_{\omega=1}^3\Gamma( N_{kl}^{\omega'}+1)}{\Gamma(\sum_{\omega=1}^3 \left( N_{kl}^{\omega'} +1)\right)}}{ \prod_{k=1}^K\prod_{l=1}^K  \frac{\prod_{\omega=1}^3\Gamma( N_{kl}^{\omega}+1)}{\Gamma(\sum_{\omega=1}^3 \left( N_{kl}^{\omega} +1)\right)} } \times 
 \frac{\prod_{k=1}^{K-1}\Gamma(n_k^{'}+1) \frac{\Gamma(K-1)}{\Gamma(N+K-1)} }{\prod_{k=1}^{K}\Gamma(n_k+1) \frac{\Gamma(K)}{\Gamma(N+K)} } \times \frac{K!}{(K-1)!}\nonumber\\
&=\frac{ \prod_{k=1}^{K+1}\prod_{l=1}^{K+1}  \frac{\prod_{\omega=1}^3\Gamma( N_{kl}^{\omega'}+1)}{\Gamma(\sum_{\omega=1}^3 \left( N_{kl}^{\omega'} +1)\right)}}{ \prod_{k=1}^K\prod_{l=1}^K  \frac{\prod_{\omega=1}^3\Gamma( N_{kl}^{\omega}+1)}{\Gamma(\sum_{\omega=1}^3 \left( N_{kl}^{\omega} +1)\right)} } \times 
 \frac{\prod_{k=1}^{K-1}\Gamma(n_k^{'}+1) }{\prod_{k=1}^K\Gamma(n_k+1)} \times \frac{K(N+K-1)}{K-1} \nonumber
\end{align}

\vspace*{2cm}

\newpage

\section{Additional results for season $2021/22$}

We reproduce the table from Table 1 in the manuscript presenting $\pi(K|\bm{y})$ for $K=1,2,3$. Here we have omitted $\pi(K=4|\bm{y})$ as this had an estimated probability of $3\times10^{-4}$. 

\begin{table}[H]
	\centering
	\begin{tabular}{l|c|c|c}
		
		$K$ & 1& 2     &  3  \\
		\hline 
		$\pi(K|\bm{y})$ & $0.0$ & $0.97$ & $0.03$ \\
	\end{tabular}
	\caption{Posterior probabilities for values of $K$}
	\label{tab:postK}
\end{table}

\begin{table}[ht]
	\centering
	\caption{Model $K = 2$: Posterior allocation probability (as a percentage) for each team} 
	\begin{tabular}{rrrrrrrrrrr}
		\toprule
		& ARS & AVL & BRE & BHA & BUR & CHE & CRY & EVE & LEE & LEI \\ 
		\midrule
		Cluster 1 & 38.29 & 0.00 & 0.02 & 0.50 & 0.00 & 89.11 & 0.03 & 0.00 & 0.00 & 0.15 \\ 
		Cluster 2 & 61.71 & 100.00 & 99.98 & 99.50 & 100.00 & 10.89 & 99.97 & 100.00 & 100.00 & 99.85  \\ 
		\toprule
		& LIV & MCI & MUN & NEW & NOR & SOU & TOT & WAT & WHU & WOL \\
		\midrule
		Cluster 1 & 100.00 & 100.00 & 1.95 & 0.04 & 0.00 & 0.01 & 74.75 & 0.00 & 0.86 & 0.53  \\
		Cluster 2  & 0.00 & 0.00 & 98.05 & 99.96 & 100.00 & 99.99 & 25.25 & 100.00 & 99.14 & 99.47 \\
		\bottomrule
	\end{tabular}
\end{table}

\begin{table}[ht]
	\centering
	\caption{Model $K = 3$: Posterior allocation probability (as a percentage) for each team} 
	\begin{tabular}{rrrrrrrrrrr}
		\toprule
		& ARS & AVL & BRE & BHA & BUR & CHE & CRY & EVE & LEE & LEI \\
		\midrule
		Cluster 1 & 27.52 & 0.27 & 0.80 & 1.25 & 0.00 & 60.68 & 0.01 & 0.00 & 0.00 & 0.36 \\ 
		Cluster 2 &  16.28 & 0.57 & 1.51 & 4.99 & 0.37 & 18.86 & 1.22 & 1.27 & 0.60 & 0.48  \\ 
		Cluster 3 & 56.21 & 99.17 & 97.69 & 93.75 & 99.63 & 20.47 & 98.76 & 98.73 & 99.40 & 99.17\\ 
		
		\toprule
		& LIV & MCI & MUN & NEW & NOR & SOU & TOT & WAT & WHU & WOL \\
		\midrule
		Cluster 1 & 64.54 & 95.47 & 0.52 & 0.33 & 0.00 & 0.12 & 56.80 & 0.00 & 1.19 & 0.61 \\
		Cluster 2  & 35.46 & 4.53 & 1.30 & 0.39 & 4.23 & 0.66 & 36.40 & 5.22 & 3.99 & 1.04  \\
		Cluster 3  & 0.00 & 0.00 & 98.18 & 99.28 & 95.77 & 99.22 & 6.80 & 94.78 & 94.81 & 98.35 \\
		\bottomrule
	\end{tabular}
\end{table}

\newpage
\section{Comparison using a Uniform prior for K}

\subsection{Details of the MCMC move type when using a Uniform prior for K}

\paragraph{Uniform prior for $K$:} We choose a discrete uniform prior for $K$ between 1 and $K_{max}$, where $K_{max}$ is again a user specified upper limit on the plausible number of blocks,

$$K \sim U(1, K_{max})\, .$$

The choice of a different prior leads to the follwing changes for the acceptante probabilities:

\subsubsection{Acceptance probabilities of the MK move}
\label{sec:MK_uniform}
\subsubsection*{Insert attempt}

Under the uniform prior, all states of $K$ have the same probability thus the ratio of the posterior density at the proposed and current states can be simplified as:
\begin{align}
\frac{\pi(\bm{z},K'|\bm{y})}{\pi(\bm{z},K|\bm{y})}&= \frac{f(\bm{y}|\bm{z})\pi(\bm{z}|K')}{f(\bm{y}|\bm{z})\pi(\bm{z}|K)}\nonumber\\
&= \frac{\pi(\bm{z}|K+1)}{\pi(\bm{z}|K)} \nonumber\\
&= \frac{ \prod_{k=1}^{K+1}\Gamma(n_{k}+1)  \frac{\Gamma((K+1)) }{\Gamma(N+(K+1))}} 
{\prod_{k=1}^K\Gamma(n_k+1) \frac{\Gamma(K)}{\Gamma(N+K)}} \ . 
\label{eq:InsMassRatio_uniform}
\end{align}

Hence, similarly to (\ref{eq:InsMassRatio_simple}), (\ref{eq:InsMassRatio_uniform}) reduces to:

\begin{align} \frac{\pi(\bm{z},K'|\bm{y})}{\pi(\bm{z},K|\bm{y})} &= \frac{ \frac{\Gamma((K+1)}{\Gamma(N+(K+1)}}{ \frac{\Gamma(K)}{\Gamma(N+K)}} \nonumber \\
&= \frac{K}{N+K}.
\end{align}

Finally, since we have symmetric proposal probabilities for $K<K_{max}$, the acceptance probability of the \textit{insert attempt} is:
\begin{equation} 
\alpha = \frac{K}{N+K} 
\end{equation}

\subsubsection*{Delete attempt}
In an analous way, the probability ratio of the probability density at the proposed and current states under a uniform prior for $K$ is:

\begin{align}
\frac{\pi(\bm{z},K'|\bm{y})}{\pi(\bm{z},K|\bm{y})}&= \frac{f(\bm{y}|\bm{z})\pi(\bm{z}|K')}{f(\bm{y}|\bm{z})\pi(\bm{z}|K)}\nonumber\\
&= \frac{\pi(\bm{z}|K-1)}{\pi(\bm{z}|K)} \nonumber\\
&= \frac{ \prod_{k=1}^{K-1}\Gamma(n_{k}+1)  \frac{\Gamma((K-1)) }{\Gamma(N+(K-1))}} 
{\prod_{k=1}^{K}\Gamma(n_k+1) \frac{\Gamma(K)}{\Gamma(N+K})} \ .  \label{eq:AddMassRatio_uniform}
\end{align}

Using the same argument as for (\ref{eq:AddMassRatio}), (\ref{eq:AddMassRatio_uniform}) reduces to 
\[\frac{\pi(\bm{z},K'|\bm{y})}{\pi(\bm{z},K|\bm{y})} = \frac{N+K-1}{K-1} \, .\]
The acceptance probability of \textit{delete} attempt when $K>1$, since this results in symmetric proposal probabilities, is:
\begin{equation}
\alpha = \frac{N+K-1}{K-1} .  
\end{equation}

\subsubsection{Acceptance probabilities for the Metropolis-within-Gibbs move}

This move and thus its associated acceptance probabilities is unchanged under the use of a different prior for $K$.

\subsubsection{Acceptance probabilities of the AE move}

\subsubsection*{Ejection attempt}
Similar to Appendix \ref{sec:MK_uniform}, the posterior probability ratio for the \textit{ejection attempt} takes the following form:
\begin{align*} 
\frac{\pi(\bm{z}',K'|Y)}{\pi(\bm{z},K|Y)} &= \frac{f(Y|\bm{z}') \pi(\bm{z}'|K')}{f(Y|\bm{z}) \pi(\bm{z}|K)}\\
&=\frac{f(Y|\bm{z}') \pi(\bm{z}'|K+1)}{f(Y|\bm{z}) \pi(\bm{z}|K)}\\
&= \frac{ \prod_{k=1}^{K+1}\prod_{l=1}^{K+1}  \frac{\prod_{\omega=1}^3\Gamma( N_{kl}^{\omega'}+1)}{\Gamma(\sum_{\omega=1}^3 \left( N_{kl}^{\omega'} +1)\right)}}{ \prod_{k=1}^K\prod_{l=1}^K  \frac{\prod_{\omega=1}^3\Gamma( N_{kl}^{\omega}+1)}{\Gamma(\sum_{\omega=1}^3 \left( N_{kl}^{\omega} +1)\right)} } \times 
\frac{\prod_{k=1}^{K+1}\Gamma(n_k^{'}+1) \frac{\Gamma(K+1)}{\Gamma(N+K+1)} }{\prod_{k=1}^K\Gamma(n_k+1) \frac{\Gamma(K)}{\Gamma(N+K)} }\\
&=\frac{ \prod_{k=1}^{K+1}\prod_{l=1}^{K+1}  \frac{\prod_{\omega=1}^3\Gamma( N_{kl}^{\omega'}+1)}{\Gamma(\sum_{\omega=1}^3 \left( N_{kl}^{\omega'} +1)\right)}}{ \prod_{k=1}^K\prod_{l=1}^K  \frac{\prod_{\omega=1}^3\Gamma( N_{kl}^{\omega}+1)}{\Gamma(\sum_{\omega=1}^3 \left( N_{kl}^{\omega} +1)\right)} } \times 
\frac{\prod_{k=1}^{K+1}\Gamma(n_k^{'}+1) }{\prod_{k=1}^K\Gamma(n_k+1)} \times \frac{K}{N+K}\, .
\end{align*}

\subsubsection*{Absorption attempt}
The posterior probability ratio is given by:
\begin{align} \frac{\pi(\bm{z}',K'|Y)}{\pi(\bm{z},K|Y)} &= \frac{f(Y|\bm{z}') \pi(\bm{z}'|K')}{f(Y|\bm{z}) \pi(\bm{z}|K)} \nonumber\\
&=\frac{f(Y|\bm{z}') \pi(\bm{z}'|K-1)}{f(Y|\bm{z}) \pi(\bm{z}|K)}\nonumber \\
&=\frac{ \prod_{k=1}^{K-1}\prod_{l=1}^{K-1}  \frac{\prod_{\omega=1}^3\Gamma( N_{kl}^{\omega'}+1)}{\Gamma(\sum_{\omega=1}^3 \left( N_{kl}^{\omega'} +1)\right)}}{ \prod_{k=1}^K\prod_{l=1}^K  \frac{\prod_{\omega=1}^3\Gamma( N_{kl}^{\omega}+1)}{\Gamma(\sum_{\omega=1}^3 \left( N_{kl}^{\omega} +1)\right)} } \times 
\frac{\prod_{k=1}^{K-1}\Gamma(n_k^{'}+1) \frac{\Gamma(K-1)}{\Gamma(N+K-1)} }{\prod_{k=1}^{K}\Gamma(n_k+1) \frac{\Gamma(K)}{\Gamma(N+K)} } \nonumber\\
&=\frac{ \prod_{k=1}^{K+1}\prod_{l=1}^{K+1}  \frac{\prod_{\omega=1}^3\Gamma( N_{kl}^{\omega'}+1)}{\Gamma(\sum_{\omega=1}^3 \left( N_{kl}^{\omega'} +1)\right)}}{ \prod_{k=1}^K\prod_{l=1}^K  \frac{\prod_{\omega=1}^3\Gamma( N_{kl}^{\omega}+1)}{\Gamma(\sum_{\omega=1}^3 \left( N_{kl}^{\omega} +1)\right)} } \times 
\frac{\prod_{k=1}^{K-1}\Gamma(n_k^{'}+1) }{\prod_{k=1}^K\Gamma(n_k+1)} \times \frac{N+K-1}{K-1}\, . \nonumber
\end{align}

\subsubsection{Comparison of posterior results}

In order to compare the use of a uniform prior for $K$ (where $K_{max}=20$) with respect to the truncated Poisson prior in the paper, we report the most relevant posterior summaries, namely the posterior probability density of $K$ (equivalent to Table 4 in the manuscript) 
and the barplot displaying the posterior estimate of the number of teams allocated to the strongest block each season (equivalent to Figure 5 in the manuscript). 

\begin{table}
	\centering
	\begin{tabular}{cc}
		
		\begin{tabular}{c|ccccc}
			\hline
			& \multicolumn{2}{c}{Number of clusters}\\
			Season & 1 & 2 & 3 & 4 & 5 \\ 
			\hline
			78/79 & 0.44 & \cellcolor{powderblue(web)}91.01 & 8.37 & 0.18 & 0.00 \\ 
			79/80 & \cellcolor{navajowhite2}93.73 & 5.77 & 0.48 & 0.02 & 0.00 \\ 
			80/81 & 12.85 & \cellcolor{powderblue(web)}85.84 & 1.29 & 0.02 & 0.00 \\ 
			81/82 &\cellcolor{navajowhite2} 93.65 & 6.23 & 0.12 & 0.00 & 0.00 \\ 
			82/83 & \cellcolor{navajowhite2}99.56 & 0.43 & 0.01 & 0.00 & 0.00 \\ 
			83/84 & \cellcolor{navajowhite2}97.82 & 2.07 & 0.11 & 0.00 & 0.00 \\ 
			84/85 & 26.39 & \cellcolor{powderblue(web)}70.83 & 2.71 & 0.07 & 0.00 \\ 
			85/86 & 0.04 & \cellcolor{powderblue(web)}99.44 & 0.52 & 0.00 & 0.00 \\ 
			86/87 & \cellcolor{navajowhite2}98.92 & 1.05 & 0.02 & 0.00 & 0.00 \\ 
			87/88 & 5.84 & \cellcolor{powderblue(web)}92.85 & 1.28 & 0.03 & 0.00 \\ 
			88/89 & \cellcolor{navajowhite2}97.99 & 2.00 & 0.01 & 0.00 & 0.00 \\ 
			89/90 & \cellcolor{navajowhite2}96.40 & 3.04 & 0.27 & 0.27 & 0.01 \\ 
			90/91 & 28.68 & \cellcolor{powderblue(web)}68.36 & 2.82 & 0.13 & 0.00 \\ 
			91/92 & \cellcolor{navajowhite2}84.32 & 15.55 & 0.13 & 0.00 & 0.00 \\ 
			92/93 & \cellcolor{navajowhite2}97.38 & 2.55 & 0.07 & 0.00 & 0.00 \\ 
			93/94 & 13.75 & \cellcolor{powderblue(web)}82.72 & 3.42 & 0.10 & 0.00 \\ 
			94/95 & 9.96 & \cellcolor{powderblue(web)}78.15 & 11.49 & 0.40 & 0.01 \\ 
			95/96 & 27.60 & \cellcolor{powderblue(web)}71.47 & 0.91 & 0.02 & 0.00 \\ 
			96/97 & \cellcolor{navajowhite2}99.17 & 0.83 & 0.00 & 0.00 & 0.00 \\ 
			97/98 & \cellcolor{navajowhite2}95.60 & 4.24 & 0.15 & 0.01 & 0.00 \\ 
			98/99 & 0.03 & \cellcolor{powderblue(web)}99.34 & 0.62 & 0.01 & 0.00 \\ 
			99/00 & 40.73 & \cellcolor{powderblue(web)}56.89 & 2.36 & 0.02 & 0.00 \\ 
			\hline
		\end{tabular} 
		& \hspace*{0.6cm} 
		\begin{tabular}{c|ccccc}
			\hline
			& \multicolumn{2}{c}{Number of clusters}\\
			Season & 1 & 2 & 3 & 4 &5 \\ 
			\hline
			00/01 & \cellcolor{navajowhite2}65.10 & 26.23 & 7.41 & 1.18 & 0.08 \\ 
			01/02 & 0.05 & \cellcolor{powderblue(web)}98.46 & 1.46 & 0.03 & 0.00 \\ 
			02/03 & 41.73 & \cellcolor{powderblue(web)}55.06 & 3.08 & 0.14 & 0.00 \\ 
			03/04 & 2.13 & \cellcolor{powderblue(web)}82.00 & 15.38 & 0.49 & 0.01 \\ 
			04/05 & 0.00 & \cellcolor{powderblue(web)}99.26 & 0.74 & 0.00 & 0.00 \\ 
			05/06 & 0.16 & \cellcolor{powderblue(web)}91.44 & 7.61 & 0.78 & 0.01 \\ 
			06/07 & 3.36 & \cellcolor{powderblue(web)}88.53 & 7.92 & 0.19 & 0.00 \\ 
			07/08 & 0.00 & \cellcolor{powderblue(web)}78.36 & 18.88 & 2.71 & 0.05 \\ 
			08/09 & 0.00 & \cellcolor{powderblue(web)}96.59 & 3.30 & 0.11 & 0.00 \\ 
			09/10 & 0.00 & \cellcolor{powderblue(web)}81.66 & 17.59 & 0.74 & 0.01 \\ 
			10/11 & \cellcolor{navajowhite2}65.80 & 33.75 & 0.43 & 0.02 & 0.00 \\ 
			11/12 & 0.56 & \cellcolor{powderblue(web)}94.38 & 4.99 & 0.07 & 0.00 \\ 
			12/13 & 0.00 & \cellcolor{powderblue(web)}98.55 & 1.44 & 0.01 & 0.00 \\ 
			13/14 & 0.00 & \cellcolor{powderblue(web)}96.06 & 3.86 & 0.08 & 0.00 \\ 
			14/15 & 3.92 & \cellcolor{powderblue(web)}83.77 & 12.03 & 0.28 & 0.00 \\ 
			15/16 & \cellcolor{navajowhite2}57.24 & 41.58 & 1.14 & 0.04 & 0.00 \\ 
			16/17 & 0.00 & \cellcolor{powderblue(web)}96.65 & 3.29 & 0.06 & 0.00 \\ 
			17/18 & 0.00 & \cellcolor{powderblue(web)}92.84 & 6.88 & 0.28 & 0.01 \\ 
			18/19 & 0.00 & \cellcolor{powderblue(web)}92.84 & 6.92 & 0.25 & 0.00 \\ 
			19/20 & 0.96 & \cellcolor{powderblue(web)}94.25 & 4.73 & 0.06 & 0.00 \\ 
			20/21 & 6.89 & \cellcolor{powderblue(web)}84.59 & 8.13 & 0.37 & 0.01 \\ 
			21/22 & 0.04 & \cellcolor{powderblue(web)}88.61 & 10.85 & 0.50 & 0.00 \\ 
			\hline
		\end{tabular}
	\end{tabular}
	
	\caption{Posterior probability of $K$, expressed as a percentage, over the last $44$ seasons, under a uniform prior assumption for $K$. The block model with highest posterior probability is coloured accordingly.}
	\label{table:allResults_uniform}
\end{table}

\begin{figure}[H]
	\centering
	\includegraphics[scale=0.33]{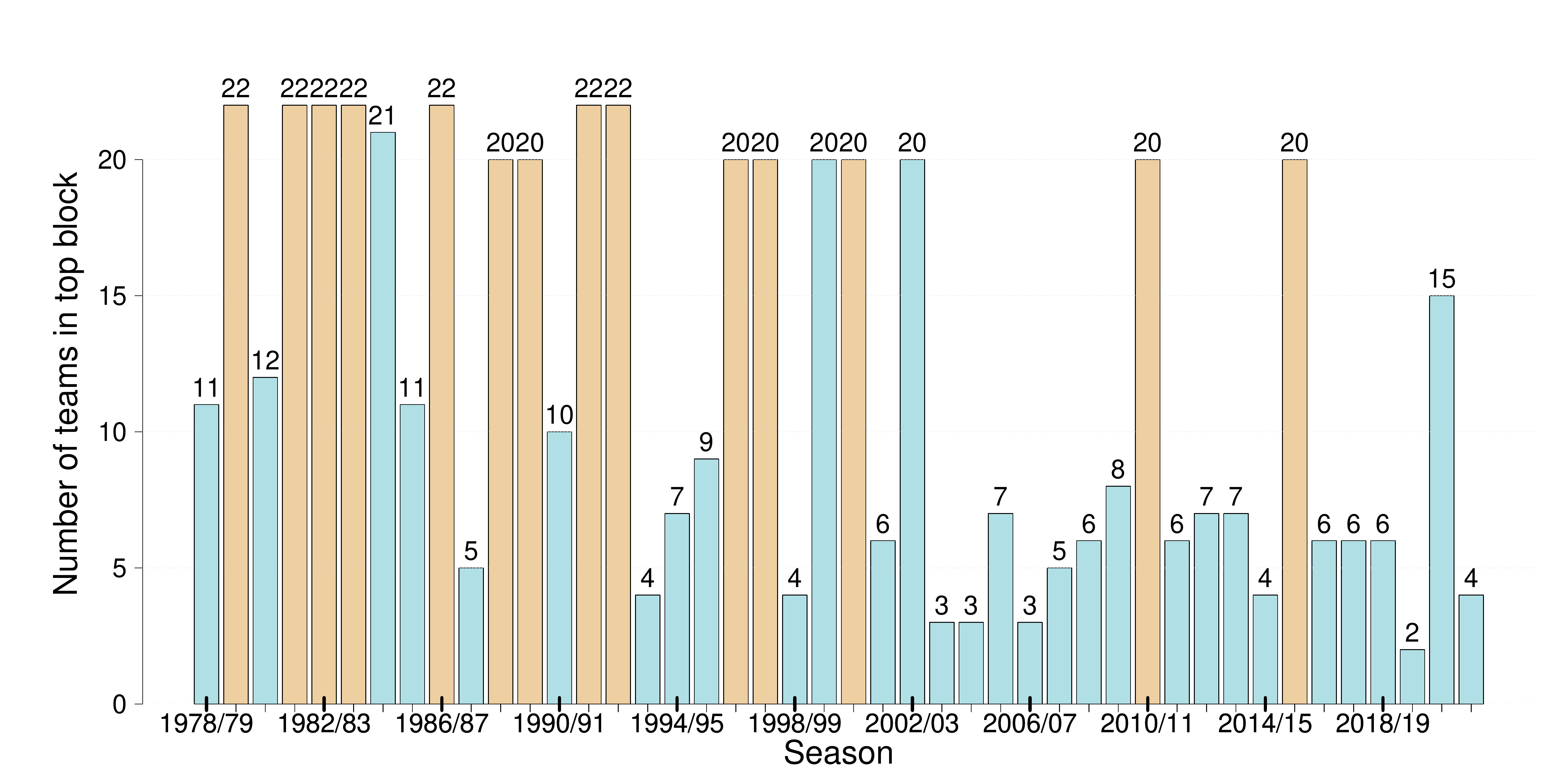}
	\caption{Barplot displaying the posterior estimate of the number of teams allocated to the strongest block each season under a uniform prior assumption for $K$. 	}
	\label{fig:topblocksize_uniform}
\end{figure}

The most substantial differences by comparing the posterior results for each season with the two priors are the following:
\begin{itemize}
	\item Season $1990/91$: 
	For this season, both posterior distributions of $K$ give stronger evidence for $K=2$. However the posterior of allocation of teams $\bm{z}$ in the top block under the uniform prior, integrating over the uncertainty in $K$, places $10$ teams in the strongest block. This constrasts with the corresponding situation for the trunctated Poisson prior where all $20$ teams are placed in the strongest block. 
	However, for the truncated Poisson prior, the posterior probability for $K =2$ is only slighly higher than for $K = 1$, which explains the difference in the posterior allocation to the top block after integrating out $K$.
	\item Season $1993/94$: The uniform prior allocates $4$ teams in the top block instead of $5$ following the truncated Poisson. Note that the posterior probabilities of $K$ under the two priors are in agreement with each other. 
	\item Season $1995/96$: There are $9$ teams in the top block according to the Uniform prior instead of $20$ following the trunctated Poisson prior. A similar comment applies as for Season $1990/91$ above.
	\item Season $1999/2000$ and $2002/03$: both priors allocate all $20$ teams in the top block but the posterior of $K$ under the Uniform prior gives highest posterior probability to $K = 2$, while for the truncated Poisson prior, the highest posterior probability corresponds to $K=1$.
	\item Season $2014/2015$: There are $4$ teams in the top block instead of $5$ truncated Poisson prior. In both cases, the highest posterior probability for $K$ corresponds to $K =2$.
\end{itemize}

We can conclude that the Uniform prior generally leads to more blocks, a posteriori, than the truncated Poisson prior, as one would expect. 
However, despite this, it does not lead to many substantial differences in the posterior allocation of teams to the strongest block, after integrating out the uncertainty in $K$. However, the truncated Poisson prior finds a methodological justification as discussed in Nobile and Fearnside (2005), essentially reflecting the fact that an SBM with $K$ blocks, there are $K!$ permutations of the block labels. For this reason, we restate our preference for the use of a truncated Poisson prior for $K$.

\newpage

\section{Team abbreviations (Season $2021/22$)}

\begin{table}[ht]
 \begin{tabular}{cc}
   ARS & Arsenal \\
   AST & Aston Villa \\
   BRE & Brentford \\
   BHA & Brighton \\
   BUR & Burnley \\ 
   CHE & Chelsea \\
   CRY & Crystal Palace \\
   EVE & Everton \\ 
   LEE & Leeds United \\
   LEI & Leicester City\\ 
   LIV & Liverpool \\ 
   MCI & Manchester City \\
   MUN & Manchester United \\ 
   NEW & Newcastle United \\ 
   NOR & Norwich City\\
   SOU & Southampton \\ 
   TOT & Tottenham Hotspur \\ 
   WAT & Watford \\ 
   WHU & West Ham United \\ 
   WOL & Wolverhampton Wanderers \\ 
 \end{tabular}

\end{table}

\end{appendices}

\end{document}